\documentclass[twocolumn]{aastex631}
\usepackage{xspace}
\usepackage{color}

\DeclareRobustCommand{\ion}[2]{%
\relax\ifmmode
\ifx\testbx\f@series
{\mathbf{#1\,\mathsc{#2}}}\else
{\mathrm{#1\,\mathsc{#2}}}\fi
\else\textup{#1\,{\mdseries\textsc{#2}}}%
\fi}

\def\arcsec{\hbox{$^{\prime\prime}$}\xspace}

\newcommand{\kms}{km\,s$^{-1}$\xspace}

\newcommand{\hst}{{\it HST}\xspace}
\newcommand{\ec}{$\eta$\,Car\xspace}

\DeclareRobustCommand{\ion}[2]{%
\relax\ifmmode
\ifx\testbx\f@series
{\mathbf{#1\,\mathsc{#2}}}\else
{\mathrm{#1\,\mathsc{#2}}}\fi
\else\textup{#1\,{\mdseries\textsc{#2}}}%
\fi}

\begin{document}
\title[excitation]{ 
 Long-term evolution in ionization of ejecta  illuminated by Eta Carinae
} 

\correspondingauthor{Damineli, Augusto}
\email{augusto.damineli@iag.usp.br}

\author[0000-0002-7978-2994]{Damineli,~Augusto}
\affiliation{Universidade de S\~ao Paulo, Instituto de Astronomia, Geof\'isica e Ci\^encias Atmosf\'ericas, \\ Rua do Mat\~ao 1226, Cidade Universit\'aria, S\~ao Paulo, Brasil}

\author[0000-0002-2806-9339]{Richardson,~Noel D.}
\affiliation{Department of Physics and Astronomy, Embry-Riddle Aeronautical University, 3700 Willow Creek Road, Prescott, AZ 86301, USA}

\author[0000-0002-0284-0578]{Navarete,~Felipe}
\affiliation{SOAR Telescope/NSF's NOIRLab, Avda Juan Cisternas 1500, 1700000, La Serena, Chile}

\author[0000-0002-6851-5380]{Gull,~Theodore.~R.}
\affiliation{Exoplanets \& Stellar Astrophysics Laboratory, NASA/Goddard Space Flight Center, Greenbelt, MD 20771, USA}

\author[0000-0002-9262-4456]{Fern\'andez-Laj\'us,~Eduardo}
\affiliation{Instituto de Astrofísica de La~Plata (CCT La~Plata - CONICET/UNLP), Argentina}

\author[0000-0002-4333-9755]{Moffat,~Anthony ~F.~J}
\affiliation{D\'epartement de Physique and Centre de Recherche en Astrophysique du Qu\'ebec (CRAQ)\\ Universit\'e de Montr\'eal, C.P. 6128, Succ. Centre-Ville, Montr\'eal, Qu\'ebec, H3C 3J7, Canada}

\author[0000-0001-5094-8017]{Hillier,~Desmond J.}
\affiliation{Department of Physics and Astronomy \& Pittsburgh Particle Physics, Astrophysics, and Cosmology Center (PITT PACC), \\ University of Pittsburgh, 3941 O'Hara Street,  Pittsburgh, PA 15260, USA}

\author[0000-0001-9754-2233]{Weigelt,~Gerd}
\affiliation{Max Planck Institute for Radio Astronomy, Auf dem H\"{u}gel 69, D-53121 Bonn, Germany}

\author[0000-0002-7762-3172]{Corcoran,~Michael~F.}
\affiliation{CRESST II \& X-ray Astrophysics Laboratory, NASA/Goddard Space Flight Center, Greenbelt, MD 20771, USA}
\affiliation{The Catholic University of America, 620 Michigan Avenue N.E., Washington, DC 20064, USA}

\begin{abstract}
Changes in the flux and spectrum of Eta Carinae since 1900 have been attributed to the evolution of the central binary by some. Others suggest evolution in the occulting ejecta. The brightness jump in the 1940s, which coincided with the appearance of narrow forbidden emission lines, may have been caused by the clearing and ionization of intervening circumstellar ejecta. The brightening changed at a slower pace up through forty years later. Here we continue earlier studies focused on the long-term showing that the forbidden line emission increased in the early 1990s with no noticeable increase in the brightness of the Homunculus. We interpret that the increase in narrow line emission is due to decreased extinction in the LOS from the central binary to the Weigelt clumps. In 2000, the central stellar core increased in brightness at a faster rate without associated changes in the Homunculus. By 2018, hundreds of narrow-line absorptions from singly-ionized metals in our LOS from Eta Carinae disappeared, thought to be caused by increased ionization of metals. These three events (1990, 2000, and 2018) are explained by the dissipation of circumstellar material within the Homunculus close to the binary. Combining these changes with the steadiness of the Homunculus and the primary winds over the past four decades indicates that circumstellar ejecta in our direction have been cleared.  
\end{abstract}


\section{Introduction}
\label{sectionintroduction}
Eta Carinae (\ec) is the best-observed example of a very massive stellar object in the local Milky Way. It became famous for its 1840s Great Eruption (GE) and the 1890s Lesser Eruption (LE), which are still not well understood \citep{davidson97, smith08}. By the 1900s, \ec\ was frequently monitored with its photometric variabilities and complex spectra which only reinforced its reputation as an unstable, unpredictable star \citep{mehner10b,lajus09,hirai21}. Doubly-ionized, forbidden lines appeared in the 1940s \citep{gaviola53} probably because of the dissipation in the circumstellar ejecta close to \ec. The Weigelt knots and the surrounding ejecta became directly bathed by the radiation of the star. The forbidden doubly ionized lines at intervals briefly disappeared with singly-ionized lines remaining. Archival analysis of spectra led to the discovery of the binary period based on the months-long low-ionization events \citep{damineli96}. \ec\ was found to be an eccentric, massive binary system \citep{damineli97,damineli00, corcoran17, grant20, strawn2023}. The hot secondary provides the high-ionization photons leading to the doubly-ionized narrow lines which disappear while the secondary star plunges into the extended, dense wind of the primary across the periastron. 

The discovery of the near knot-like Weigelt clumps \citep{weigelt86, hofmann88},  projected to be $\sim$\,0\farcs3 from \ec, soon were understood to be slowly moving clumps of ejecta. Spectroscopy of the Weigelt clumps, resolved from \ec, demanded sub-arcsec spatial resolution, accessible only by \hst. \citet{davidson95} demonstrated that Weigelt D and B were sources of the narrow high-ionization forbidden lines that offered a means to monitor the central binary and to model the luminosity and temperature of the secondary star \citep{verner02, verner05, Mehner2010, Teodoro20}. 
 
The apparent long-term spectral evolution of \ec\ had been attributed to intrinsic variations, such as a decrease in the mass-loss rate of the primary star or evolution of a latitudinal-dependent stellar wind \citep{mehner15, Smith_2003}. However, intrinsic wind changes conflict with observational results, such as $a)$ the stable, long-term, near-infrared and mid-infrared luminosity from 1968 to 2018 \citep{mehner19}; $b)$ the steadily repeating X-ray light curve over multiple binary orbits \citep{corcoran17, espinoza21}; $c)$ the constancy of the dust-scattered spectrum from the Homunculus \citep{hillier92,damineli21}, d) the constancy of the reflected continuum flux in Weigelt~D in a period when the star brightened \citep{mehner12} and $e)$ the repeatability of the narrow line fluxes from the ejecta. Regarding the Homunculus brightness, \citet{smith17} examined two images taken by \hst in 1995 and 1998 and found a brightness difference of 34\,\%. In the long term, however, the nebular brightness is flat, with excursions up to 0.3\,mag from the average 5.6\.mag in the F550M filter of the ACS/HRC \hst camera. See Fig.\,5 and Table A1 of \citet{damineli19}

\cite{hillier92} first suggested that an occulter must be in our LOS to \ec. Its impact on the photometric and spectroscopic evolution of \ec was further discussed by \citet{damineli19, damineli21, damineli23} and \citet{pickett2022}. In addition, \citet{gull23} showed that,  between 2004 and 2018, many singly-ionized absorption lines disappeared across an extended portion of the Homunculus
(within $\pm$\,1\farcs5 of the central stellar core), indicating that the dissipating occulter extended well beyond our LOS to \ec.

 \citet{humphreys08} reported the spectroscopic and photometric evolution of \ec, showing that the star was photometrically stable in the period from 1900 to 1940. The spectrum also remained nearly constant over that period, displaying only low excitation lines like \ion{H}{i}, \ion{Fe}{ii} and [\ion{Fe}{ii}].
In the period 1940--53 there was an episode of brightening, which was in coincidence with the appearance of \ion{He}{i}  and [\ion{Fe}{iii}] lines. The object slowly brightened up to the 1990s, the [\ion{Fe}{iii}]/[\ion{Fe}{ii}] ratio increased, while the \ion{He}{i} remained at a roughly constant level. To understand the evolution of \ec recently, two main questions need to be properly addressed: how have the spectral and photometric variations evolved in the last three decades? What caused the remarkable events that occurred around the 1940s and 1990s? 

To obtain more insight on this subject, we collected about 500 high-resolution spectra covering the last three decades (see, e.g., \citealt{damineli21, damineli23}). We used them to analyze the evolution of the degree of ionization of the ejecta clumps by measuring the peaks of the narrow lines \ion{Fe}{ii}\,$\lambda4631$, [\ion{Fe}{ii}]\,$\lambda4641$, and [\ion{Fe}{iii}]\,$\lambda4659$\footnote{We use the same line identifications as in \citet{Teodoro20}. All wavelengths are in vacuum.}. The last line is sensitive to photoionization, probing the amount of radiation from the secondary star received by the ejecta while the first two lines are largely supported by the primary stellar UV shortward of Lyman~$\alpha$. The electron density of the clumps is sufficiently high (log,N$_e$\,$\sim$\,7) to cause the [\ion{Fe}{ii}]\,$\lambda4641$ emission to be dominated by collisional processes and almost insensitive to the radiation field \citep{Mehner2010, Teodoro20}. 

The reader will note that the ionization potentials (IPs) of Fe to Fe$^+$ is 7.90 eV and Fe$+$ to Fe$^{++}$ is 16.18 eV. With H to H$^+$ at 13.6 eV, the [\ion{Fe}{ii}] and [\ion{Fe}{iii}] discriminate between the presence of radiation above and below the ionization of hydrogen.

In this scenario, the excitation of Weigelt B and D was discussed by \citet{verner02, verner05}. \cite{Gull16} described spatially resolved maps of [\ion{Fe}{ii}]~$\lambda$4815 and [\ion{Fe}{iii}]~$\lambda$4659 in detail.  \cite{Mehner2010} discussed the variability of the [\ion{Fe}{iii}~$\lambda$4659 lines for Weigelt D across  orbital cycle 10. \cite{Gull16} and \cite{Teodoro20} discussed the changes in the mapped forbidden lines across cycle 12. There were a few intrinsic cycle-to-cycle changes in the line emission, especially the disappearance in Weigelt B between cycle 10 and cycle 12 \citep{Gull16}, but without a substantial impact on the narrow-line emission fluxes. The [\ion{Fe}{iii}]\,$\lambda4659$ line changes dramatically across both orbital cycles, complicating the study of its long-term evolution without accurate comparisons at the same phases in different cycles, which motivated us to conduct the present work. 

The trade-off from monitoring with ground-based spectrophotometry versus \hst/STIS is losing all spatial- and velocity-field details to focus on the narrow line peaks with dense time sampling over many orbital cycles.  We chose to use frequent high-resolution, ground-based spectroscopy to monitor the ionization of all ejecta clumps over the last six orbits, based on the fluxes of the \ion{Fe}{ii}\,$\lambda4631$, [\ion{Fe}{iii}]\,$\lambda4641$ and  [\ion{Fe}{ii}]\,$\lambda4659$ lines. 

 The selection of these three iron lines was motivated by their wavelength proximity, sharing a common stellar continuum, and the fact that two of them - [\ion{Fe}{iii}]\,$\lambda4659$ and [\ion{Fe}{ii}]\,$\lambda4641$  - have been reported since the 1940s and their ratio is sensitive to the degree of ionization of the nebula.

\section{Data: observations and reduction}
\label{sectionobservations}
 
\subsection{Absolute flux of the forbidden emission structure with the \hst/STIS}
\label{subsection_STIS}

\begin{figure}
    \centering
    \includegraphics[width=8.5cm]{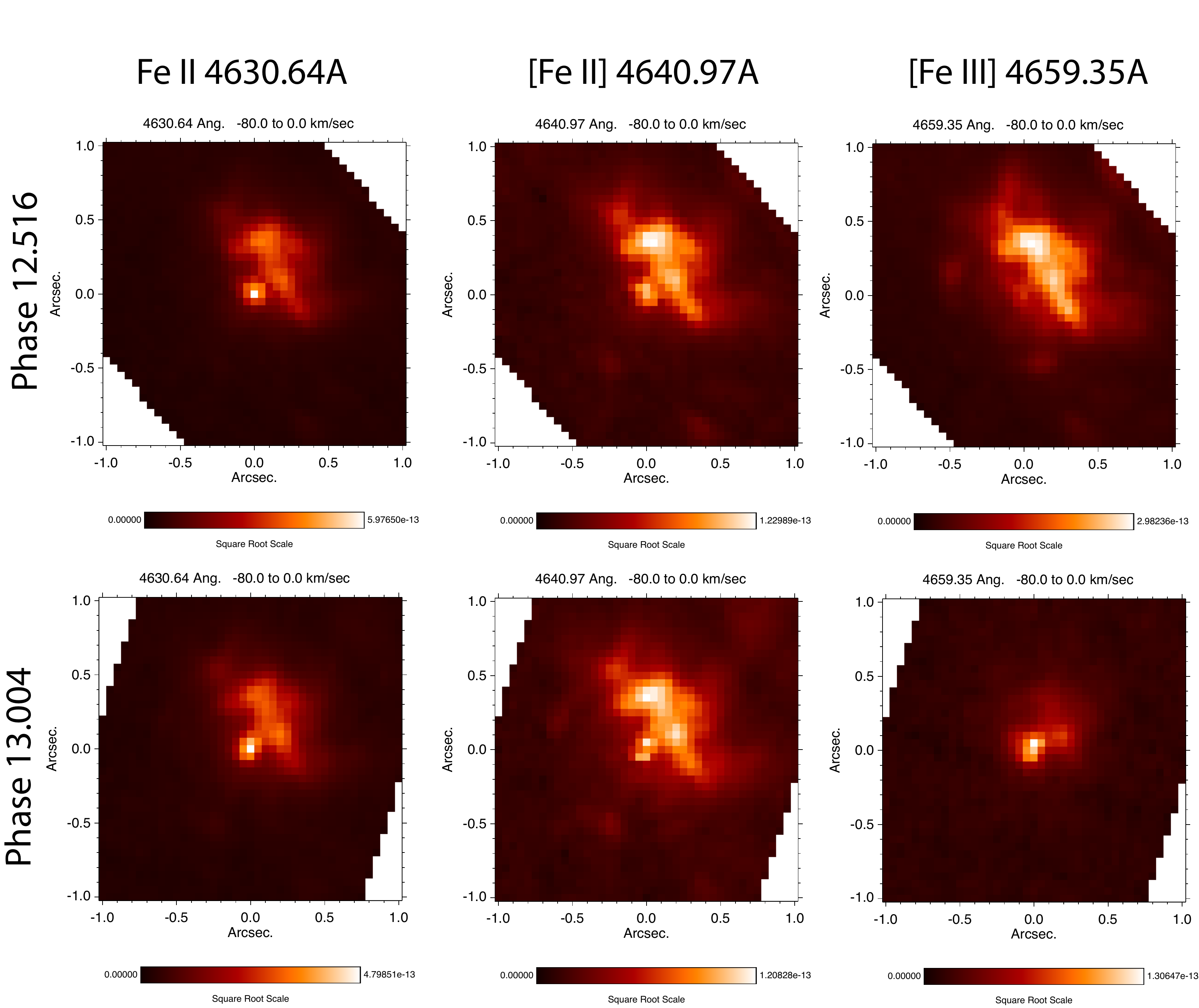}
    \caption{\hst/STIS mappings of the emission lines at the high-ionization state, 12.516, and at the low-ionization state, 13.003. The \ion{Fe}{ii}~$\lambda$4631 and the [\ion{Fe}{ii}]~$\lambda$4641 structures and intensities change minimally from high-ionization to low-ionization states. The [\ion{Fe}{iii}]~$\lambda$4659 fluxes drop during the periastron passage as \ec-B is immersed within the extended wind of \ec-A. The residual flux in the [\ion{Fe}{iii}]~4659\AA\ map at 13.004 is from [\ion{Co}{ii}]~4658.94$\AA$ and [\ion{Fe}{ii}]~4658.29$\AA$. The angular resolution is 0\farcs1 and the velocity is centered at $-$40 \kms, the velocity of the Weigelt clumps with a width of 80 \kms. The intensity display  is sqrt(flux).} 
\label{mappings}
\end{figure}

Diffraction-limited mappings of selected emission lines had been obtained using the  {\it Hubble Space Telescope}\footnote{Based on archived observations made with the NASA/ESA Hubble Space Telescope, obtained at the Space Telescope Science Institute, which is operated by the Association of Universities for Research in Astronomy, Inc., under NASA contract NAS5-26555.}  /Space Telescope Imaging Spectrograph (\hst/STIS) across one cycle of the \ec\ 5.53-year period. As \cite{Gull16} addressed changes in the structures of [\ion{Fe}{ii}]\,$\lambda$4815 compared to [\ion{Fe}{iii}]$\lambda4659$, we revisited the data to directly compare [\ion{Fe}{ii}]\,$\lambda$4641 with [\ion{Fe}{iii}]\,$\lambda$4659.

The mappings of \ion{Fe}{ii}\,$\lambda$4631, [\ion{Fe}{ii}]\,$\lambda$4641 and [\ion{Fe}{iii}]\,$\lambda$4659 are compared in Figure\,\ref{mappings} at the high-ionization state, 12.516, and at the low-ionization state, 13.003. The structure of \ion{Fe}{ii} and [\ion{Fe}{ii}] changed little between the high- and low-ionization state, but the [\ion{Fe}{iii}] drops dramatically by the low-ionization state. Further examination of the individual spectra that were used to build the map of [\ion{Fe}{iii}] and with comparison to the spectral atlas published by \cite{Zethson12} strongly suggests that the weak emission seen in [\ion{Fe}{iii}]\,$\lambda$4659 in the low-ionization state originates from two weak lines of [\ion{Fe}{ii}] at 4658.29\AA\ and [\ion{Co}{ii}] at 4658.94\AA. i.e. likely the [\ion{Fe}{iii}]\,$\lambda$4659 is completely absent during the low-ionization state, consistent with the ionizing photons necessary for Fe$^{++}$ originating from the hot secondary, \ec-B, which is enshrouded by the extended primary wind during periastron passage.

\begin{figure}
    \includegraphics[width=\linewidth, angle=0, viewport=55bp 0bp 690bp 520bp, clip]{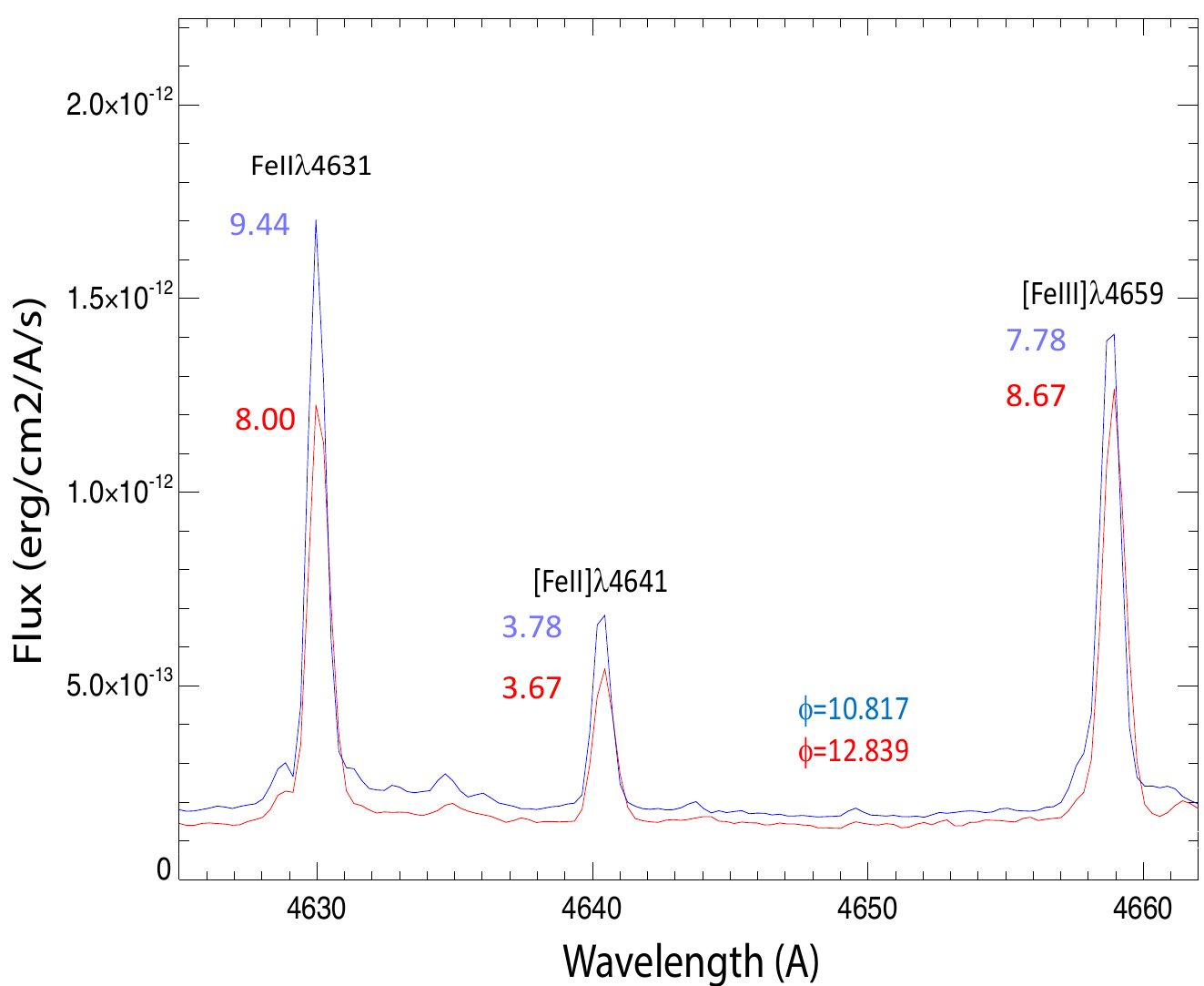}
    \caption{Spectra extracted from {\it \hst/STIS} observations centered upon Weigelt D at phases 10.817 and 12.839 demonstrate minimal changes in the interval between cycles 10 and 12, consistent with ground-based spectrophotometry in Fig.~\ref{Fig4}. Numbers near the peaks indicate continuum-normalized intensities. They are larger than in ground-based spectra because of the lower continuum flux at better spatial resolution. }
\label{2cycle}
\end{figure}

Since absolute fluxes are precious information, we searched the \hst/STIS database for visits separated by one or more cycles at the same orbital phase. We found just a pair of spectra centered on the Weigelt D knot, covering the three iron lines that we are focusing on - see Figure~\ref{2cycle}. The STIS spectra were taken with the slit width $\sim$\,0\farcs1 and extracted in a window $\sim$\,0\farcs125 long. The slit PA orientation was different for the two visits, which may cause an uncertainty of $\sim$\,10\% in the absolute flux. The two STIS spectra are just for a sanity check of our ground-based observations, which are subject to more complex factors, such as $a)$ knot D is  $\sim$\,70\% stronger than C at this phase \citep{Teodoro20}, $b)$ ground-based observations correspond to even larger line fluxes because a larger area is sampled due to the slit width, $d)$ ground-based spectra sample clumps with different physical conditions, resulting in different line ratios as compared to the single knot D observed by STIS.

\subsection{ High-spectral resolution ground-based spectroscopy from cycle-to-cycle}
\label{subsection_groundbased}

Using multiple ground-based spectrographs we gathered spectra with a signal-to-noise ratio of 100--500 and spectral resolving power ranging from R\,$\sim$90\,000 to 22,000 (see Table\,\ref{table_observat} for the origin of the spectra). Many of these spectra have been used in several recent papers by our group which details the reductions of the data \citep[see, e.g.,][]{richardson10,richardson15,richardson16,teodoro16,damineli08a,damineli19,damineli21,damineli23,pickett2022,strawn2023} The spectra were normalized by using a linear fit through the pseudo continuum in the range 4550--4750\,\AA. Even though faint lines present in these spectral windows could introduce a source of error in the normalized intensity of the lines, the main source of scattering in the line intensities is due to different spectral resolutions and differences in seeing conditions. The signal-to-noise exhibits a more significant impact for the line [\ion{Fe}{ii}]\,$\lambda4641$, which is relatively faint, than for the other two stronger lines, except for [\ion{Fe}{iii}]\,$\lambda4659$ across the low-excitation periastron passages.  

\begin{table}
	\centering									
	\caption{Observatories}
	\label{table_observat}
	\begin{tabular}{llll} 
    \hline							
Observ.	& Telescope & R &number	\\
 & aperture &  &measur. \\
\hline
LCOGT/NRES	& 1\,m    &48\,k& 212\\
CTIO/CHIRON	& 1.5\,m& 80\,k&196\\
ESO/FEROS	& 1.5\&2.2\,m&48\,k& 109\\
LNA/OPD     & 1.6\,m&22\,k&47\\
ESO/Hexapod	& 1.5\,m& 48\,k&3\\
ESO/UVES	& 8\,m& 90\,k&15\\
Mount Strom. Obs.& 1\,m& 48\,k&7\\
B.~Heatcote & 0.28\,m &16\,k&1\\
\hline
\end{tabular}
\end{table}

Differences in spectral resolution can impact the intensities since we are measuring narrow lines, especially \ion{Fe}{ii}\,$\lambda4631$ which is the narrowest with fwhm\,$=$53\,\kms, as compared to 58\,\kms for [\ion{Fe}{iii}]\,$\lambda4641$ and 84\,\kms for [\ion{Fe}{iii}]\,$\lambda4659$.

Most of the line and continuum flux originates in a region inside a $\sim$\,0\farcs5 radius. The seeing smears out the image to $\sim$1--3\arcsec, so there is not much difference between slit or fiber-fed spectrographs. An exception was noticed in a few of the UVES spectra which used a (0\farcs2  aperture) under excellent seeing conditions. The scatter cancels out very nicely through line ratios. 

The decreasing strength in the emission lines (see Fig. 3 in \citealt{mehner15}) is not intrinsic, but due to the contrast between the forming region (Weigelt clumps) and the central star brightness which increases as the intervening occulter dissipates \citep{damineli21}. But this dominant effect masks long-term intrinsic evolution. The \ion{Fe}{ii}\,$\lambda4631$ was the strongest in 1997.0 but in 2023.6, it had its peak intensity lower than that of [\ion{Fe}{iii}]\,$\lambda4659$. Labels in black indicate the peaks of the line intensities. The lines were normalized to the continuum flux and displaced vertically for clarity. Phase is calculated using P\,$=$\,2022.7\,days and $\phi$\,$=$\,14.0 is for 2020.2.

\begin{figure}[ht]
\centering
\includegraphics[width=\linewidth, angle=0, viewport=190bp 50bp 750bp 520bp, clip]{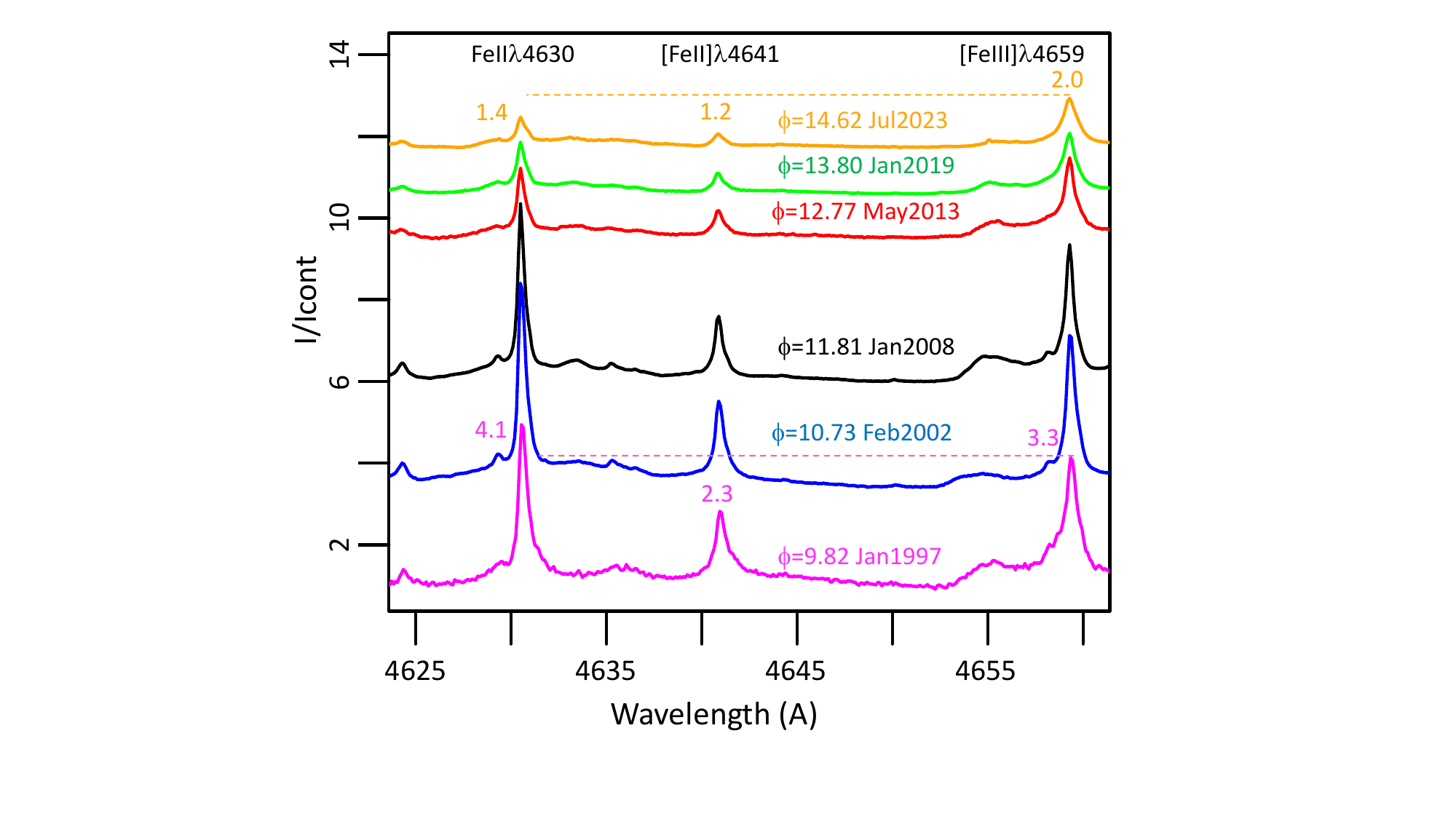}
\caption{ Ground-based evolution of continuum-normalized line intensity of three narrow iron lines. The increase in the continuum flux of the whole object (Homunculus + stellar core) is the cause of the general decrease with time of the normalized line intensities. As shown in Figure \ref{2cycle}, the flux in the Weigelt knots is reasonably constant. The evolution of the line strengths is exemplified by the numbers near the line peaks.}
\label{Fig3}
\end{figure}

We linked the modern records of the [\ion{Fe}{iii}]\,$\lambda4659$\,$/$\,[\ion{Fe}{ii}]\,$\lambda4641$ ratio to six spectra recorded on photographic plates and one spectrophotometric scan recorded before the 1990s. Photographic plates have a logarithmic response on top of an unknown background threshold.   The uncertainty in correcting photographic plate responses was challenging for the six photographic spectra - Table~\ref{table_fe3fe2}. 

 For the spectra reported by \citet{humphreys08} and \citet{zanella1984}, we measured the line intensities in the plotted scan and applied a logarithmic correction to derive the line ratio.The measurement for the 1938 ratio - Figure 10 in \citep{humphreys08} - is very uncertain as the [\ion{Fe}{iii}]\,$\lambda4659$ is exceedingly faint. However, the identification of [\ion{Fe}{iii}]\,$\lambda4659$  seems to be correct because [\ion{Fe}{iii}\,$\lambda4702$], although much fainter is clearly present. We adopted a generous upper limit of 0.3 for this ratio. Gaviola's spectra indicated a level almost three times higher just one cycle ahead at the same phase. We used the reported intensities of the spectra presented by \citet{gaviola53} and \citet{thackeray53}. 

The measure of line intensities by \citet{aller66} was calibrated with photoelectric scans. The spectrum reported by \citet{hillier92} was recorded with a CCD, and our measurement is very close to those reported in Table~2 of \citealt{humphreys08}. We adopt the values of those authors for the four measurements in common.

Our experiments with modern spectra indicate that this ratio is affected by the spectral resolution although by a reasonable, predictable amount.

The V$-$band light curve was collected from many sources - especially \citet{smith11} and \citet{frew04} - and assembled by doing small zero-point shifts to bring the magnitudes \citep{lajus09, damineli19} to the same scale as the La Plata monitoring campaign (http://etacar.fcaglp.unlp.edu.ar/EtaCar/).
 Most of these measurements were extracted inside a circular-aperture radius of 12\, arcsec \citep{lajus09} in a differential-magnitude scheme,  using HDE303308 as a comparison and resulting in an accuracy of $\sim$\,0.01 mag. In addition to the V$-$band photometry, we used the TG bandpass reported in AAVSO (https://www.aavso.org/), which conforms nicely with the V$-$band after a small shift in the zero-point.

All the facilities are listed in Table~\ref{table_observat} and the corresponding observational programs are provided in the Acknowledgements. 

\section{Results}
\label{sectionresults}

\begin{figure*}[ht!]
\centering
{\includegraphics[width=1\linewidth,angle=0,viewport=240bp 140bp 720bp 375bp, clip]{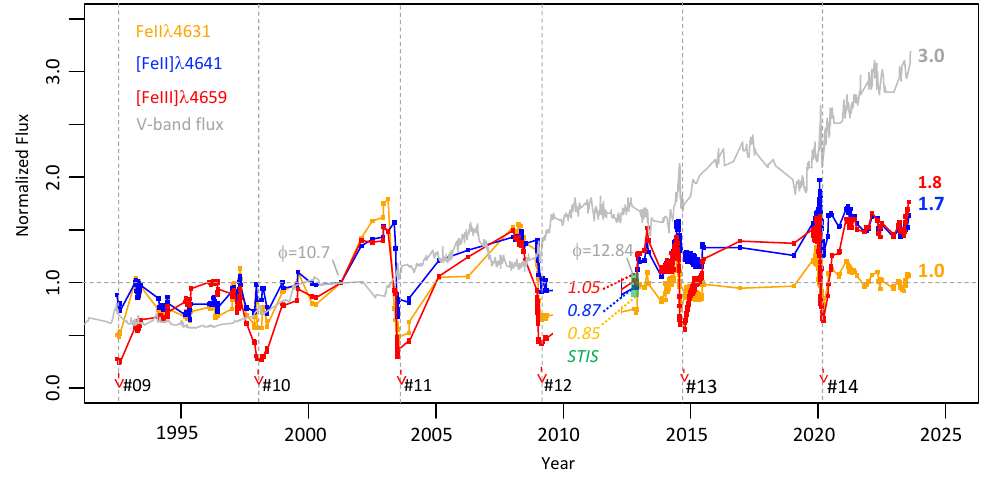}}
\caption{Time series of the three iron lines in the last 30 years. Line intensities are multiplied by the V$-$band flux (gray line). All fluxes are normalized to the unit value in 2002.1, ($\phi$\,$=$\,10.7) - gray label. This phase and $\phi$\,$=$\,12.8 correspond approximately to dates when the STIS spectra in Fig.~\ref{2cycle} were taken. STIS fluxes at $\phi$\,$=$\,12.839 relative $\phi$\,$=$\,10.817 are indicated in italics in cycle 12 - small green rectangle. The STIS fluxes over these three cycles are $\sim$\,10\% higher than those taken with ground-based spectra. Labels to the right of each time series indicate the value in 2023.6.  Labels at the bottom, placed near the vertical dashed lines, indicate the running number of low-excitation events. }
\label{Fig4}
\end{figure*}

\subsection{Line-flux evolution from cycle to cycle}
\label{subsectionintensities}

The line-intensity fluxes of the three iron lines observed by STIS shown in Figure~\,\ref{2cycle} (blue for $\phi$\,$=$\,10.817 and red for $\phi$\,=\,12.839) are much more prominent than in ground-based spectra (Figure~\ref{Fig3}) at the same phases because STIS probes much fainter continuum flux at better spatial resolution. Except for the \ion{Fe}{ii} line, which seems roughly fainter at $\phi$\,$=$\,12.839, the two forbidden lines remained at the same flux within the uncertainties. The [\ion{Fe}{iii}]\,$\lambda4659$\,$/$\,[\ion{Fe}{ii}]\,$\lambda4641$ flux ratio remained reasonably constant for the same phase separated by three cycles. The reader should note that Figure \ref{2cycle} presents the spectrum of Weigelt D only, not of the entire group of emission clumps.

For ground-based observations, the most evident evolutionary pattern of the continuum-normalized, narrow iron emission lines is a generally strong fading with time, as shown in Fig.\,\ref{Fig3}. The dates of the spectra were chosen to be in high-excitation phases ($\phi$\,$\sim$\,0.6--0.8) when intensity variations along the cycle are relatively small. These lines suffer a significant fading factor over a long time scale because $1)$ the stellar continuum is increasing as the extinction of the occulter along our LOS to the stellar core is decreasing, and $2)$ the clumps are located outside of that region of extinction, which is consistent with the decreasing intensity observed for the  \ion{Fe}{ii}\,$\lambda8610$ line \citep{damineli23}. In addition to the general weakening of the line intensities, the ratio between the line peaks of [\ion{Fe}{iii}]\,$\lambda4659$ to \ion{Fe}{ii}\,$\lambda4631$ changed from 0.8 to 1.4, suggesting that the excitation degree in the ejecta clumps increased with time. Note that this is not related to the contamination of the central source since the continuum is the same for the three lines. If this effect is real, the ejecta clumps are receiving more high-ionization flux now than in the 1940s. Is this due to a decreasing mass loss rate of the primary star or to the partial dissipation of ejecta along the LOS from the Weigelt clumps to the secondary star? In the next sub-sections, we will try to answer these questions by investigating the time series of line intensities.

\subsection{Time series of line intensities}
\label{subsection_timeseris}

Long-term spectral time series encompassing multiple orbital cycles is a key ingredient to obtaining a more comprehensive understanding of the evolution of spectral lines. Since the most prominent component of the lines originated in the ejecta clumps is narrow, high spectral resolving power (R\,$>$\,20,000) is necessary. The second most important factor in a long-time series is to keep the spectral resolution as constant as possible. The spatial resolution is not a key factor in ground-based observations to intercompare line intensities from different spectrographs since a slit or a fiber sample the inner 1--3\farcs5 diameter around the source, where most of the continuum is formed. Most of the nebula does not contribute to the stellar continuum. The full list of the spectral measurements analyzed in this section is presented in Table~\ref{table_lineintensity}.

\begin{table}							
\centering							
\caption{Line peak intensity ratio of [Fe\,{\sc{iii}}]$\lambda 4659$/[Fe\,{\sc{ii}}]$\lambda 4641$, measured at different epochs and orbital phases of the $\eta$\,Car system. }							
\label{table_fe3fe2}							
\begin{tabular}{lllll} 							
\hline							
Year	&	Phase	&	Ratio &	comm. & Reference	\\
\hline							
1938	&	-0.85	&	$<$\,0.3 & -- &	H08\\
1944	&	0.24	&	0.8	 &--&	G53	\\
1951.22	&	1.54	&	0.7	 &--&	T53	\\
1961.25	&	3.35	&	1.00	 &--&	A66	\\
1974.08	&	5.67	&	1.05  &--&    Z84\\
1983.13	&	7.30	&	1.00	 &--&	Z84\\
1986.22	&	7.88	&	1.15	 &--&	H92\\
1994.15	&	9.31	& 1.30  &R=5k&	ESO/FEROS\\
1996.35	&	9.71	& 1.65	&R=5k&	ESO/FEROS	\\
1999.58	&	10.29	& 1.32	&R=5k&	ESO/FEROS	\\
2002.5	&	10.82	& 1.40	&R=5k&	ESO/FEROS	\\
2005.19	&	11.29	&1.33	&R=5k&	ESO/FEROS	\\
2008.02	&	11.81   &1.43	&R=5k&	ESO/UVES	\\
2013.30	&	12.77	&1.49	&R=5k&	CTIO/CHIRON	\\
2016.94	&	13.42	&1.36   &R=5k&	ESO/FEROS\\
2019.05	&	13.82	&1.54	&R=5k&	LCOGT/NRES	\\
2021.24	&	14.31	&1.49	&R=5k&	LCOGT/NRES	\\
2023.57	&	14.62	&1.44	&R=5k&	LCOGT/NRES	\\
\hline
2002.6	&	10.82	&2.06   &Weigelt D&HST/STIS	\\
2013.8	&	12.84	&2.36   &Weigelt D&HST/STIS	\\
\hline							
\end{tabular}
\\
\textbf{Notes:}
References: H92=\citet{hillier92}; H08=\citet{humphreys08}; G53=\citet{gaviola53}; T53=\citet{thackeray53}; Z84=\citet{zanella1984};
R\,=\,5\,000 - degraded resolution to compare with 1938-86 ratio.
\end{table}

\begin{figure*}[ht!]
\centering
{\includegraphics[width=1\linewidth,angle=0,viewport= 240bp 100bp 720bp 410bp, clip]{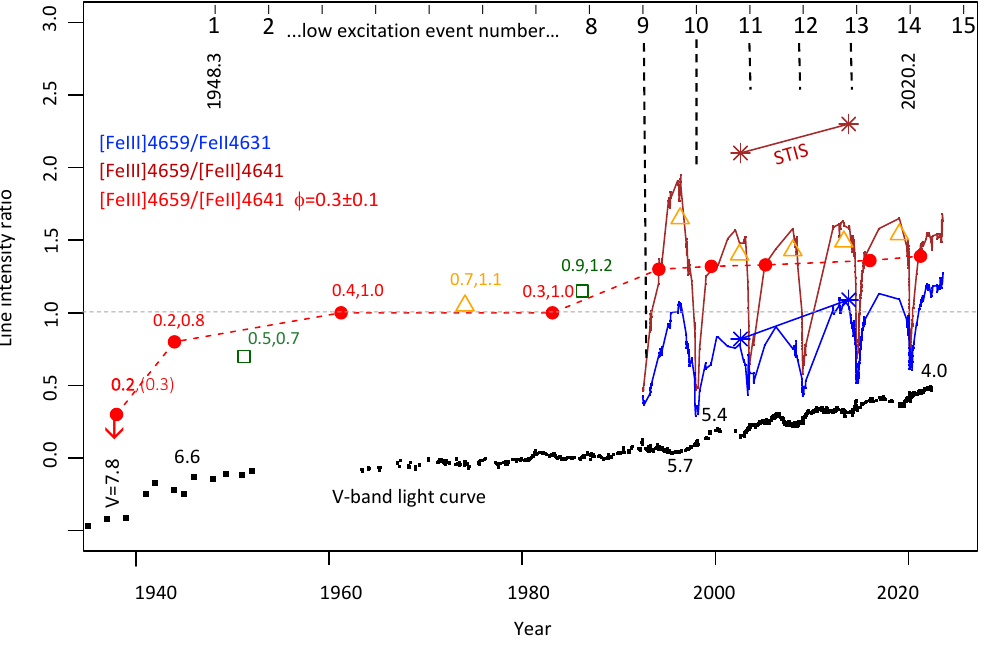}}
\caption{ Iron line ratios estimated from the continuum-normalized spectra. Ground-based measures of the [\ion{Fe}{iii}]\,$\lambda4659$\,$/$\,[\ion{Fe}{ii}]\,$\lambda4641$ ratio are represented by small brown dots reported in Table \ref{table_lineintensity}. 
The blue line represents the flux ratio [\ion{Fe}{iii}]\,$\lambda4659$\,$/$\, \ion{Fe}{ii}\,$\lambda4631$. Brown and blue asterisks represent STIS measures of the two ratios. The red-filled circles are the ratios after degrading the spectra to R\,$=$\,5\,000 - Table~\ref{table_fe3fe2}. The focus on $\phi$\,$=$\,0.3 is to link them with the historical measurements taken at $\phi$\,$=$\,0.3\,$\pm$\,0.1 also reported in Table~\ref{table_fe3fe2}. Orange triangles indicate line-flux ratios for high excitation phases ($\phi$\,$=$\,0.7-0.8) from spectra degraded to R\,$=$\,5\,000.
 In the upper axis, the times of the low-excitation events are marked (01 is for the first one observed by \citet{gaviola53} in 1948.3). The V$-$band light curve is displayed at the bottom (black squares), and the black numbers close to it indicate the magnitude of the whole object (core+Homunculus).}
\label{Fig5}
\end{figure*}

 {The best strategy to follow the long-term evolution of these three lines -  [\ion{Fe}{iii}]\,$\lambda4659$, [\ion{Fe}{ii}]\,$\lambda4641$, and \ion{Fe}{ii}\,$\lambda4631$ - consists of the conversion of their intensities relative to the stellar continuum to relative fluxes, as shown in Fig.~\ref{Fig4}.} The conversion is done by multiplying their intensities by the $V$-band flux light curve (gray line) observed from the ground, which encompasses the entire object (core plus the Homunculus). The absolute flux accuracy is irrelevant since we are dealing with flux ratios. 

On top of the secular brightening of the object, some oscillations became more pronounced as the stellar core became dominant over the nebula in the latest cycles. The oscillation amplitude is $\sim$\,0.4\, mag due to the variation of the projected shape of the primary's to our LOS, distorted by the cavity produced by the wind-wind collision when it goes around the orbit. The amplitude of the light-curve oscillation isolating just the stellar core is almost twice the whole object because the Homunculus adds a constant light corresponding to V$_{H}$\,$\sim$\,5.5~mag (see Fig.~16 from \citealt{damineli19}).

The spectral-line time-series and the V$-$band light-curve were normalized to their values at 2002.1 ($\phi$\,=\,10.7) - Fig.~\ref{Fig4}. This choice is arbitrary and was selected to be as close as possible to the phase of the spectrum taken with STIS at $\phi$\,$=$\,10.817. A second spectrum was taken in the same position (Weigelt~D) two cycles later, at phase $\phi$\,$=$\,12.839 (Fig.~\ref{2cycle}).
The absolute fluxes measured in Weigelt~D show almost no temporal evolution between the two epochs - see Fig.~\ref{Fig4}. Their relative fluxes are $\sim$\,10\,\% higher than the corresponding ground-based observations. This demonstrates that the spectrophotometry from the ground recovers the line flux of the ejecta clumps with fairly good accuracy when compared to the analysis of STIS spectroscopy. The two dates when the STIS spectra were taken are shown as gray numbers and marked with a small green rectangle in cycle 12 in Fig.~\ref{Fig4}.

The behavior of the [\ion{Fe}{iii}]\,$\lambda4659$ line along the orbital cycles presented in Fig.~\ref{Fig4} is in agreement with that discussed by \citet{Mehner2010} for cycle 10, and by \citet{Teodoro20} for cycle 12. There is a sharp minimum at phase zero (probably reaching zero intensity) and fast recovery to the high-excitation phases leading to a pronounced peak around phase 0.9 (about three months before phase zero)\footnote{The red downward arrows in Figure\,\ref{Fig4} at the zero phases indicate our inability from the ground to follow the fading of [\ion{Fe}{iii}]\,$\lambda4659$ below the stellar continuum. Its real intensity can be up to ten times fainter in the low-excitation phase than in mid-cycle \citep{Mehner2010, Teodoro20}}.
The situation is similar for the other two iron lines, but their minima do not go to zero intensity, and the mid-cycle maximum is less pronounced. The pre-minimum peak at the end of the cycles is not present in the ratio curve [\ion{Fe}{iii}]\,/\,[\ion{Fe}{ii}] of Fig.\,\ref{Fig3}. {This is unexpected since [\ion{Fe}{iii}] is sensitive to the ionization flux, but [\ion{Fe}{ii}] is not (see, e.g., the discussion for the Weigelt~D knot in \citealt{Teodoro20}).

The $V$-band flux has increased by a factor of three since 2001 (gray light curve in Figure~\ref{Fig4}). This flux increase is caused by the dissipation of the occulter along our LOS to the stellar core. The fact that the [\ion{Fe}{iii}] flux also increased indicates that more radiation from the ionizing secondary is reaching the ejecta clumps. Moreover, \citet{gull23} reported the disappearance of extended narrow absorption structures in yet another direction during the last orbital cycle. These three observational facts indicate that the dissipation process impacts different circumstellar regions at different times.

\subsection{ Intensity ratio of [\ion{Fe}{iii}] to [\ion{Fe}{ii}] and to \ion{Fe}{ii} lines}
\label{subsection_ratio}

 Figure~\ref{Fig5} presents the line intensity ratio curve derived from Table~\ref{table_lineintensity} - blue and brown curves. The Fe$^{++}$ to Fe$^{+}$ ratios are much smoother than the time series presenting the same data as flux intensities for each line. The plot shows a remarkable variation caused by the orbital motion of the central binary, which modulates the ionizing flux impacting on the ejecta clumps. The Fe$^{++}$ line tracks the flux from the secondary star - which plunges into the primary's wind inner region at periastron - and that of Fe$^{+}$ tracks the primary's flux, which is less subject to the orbital modulation effects. The blue line represents measurements of the ratio [\ion{Fe}{iii}]\,$\lambda4659$\,$/$\,\ion{Fe}{ii}\,$\lambda4631$. On top of the orbital modulation, this ratio shows a general long-term increase of the orbital maxima, indicating that the ionization level in the Weigelt clumps has been increasing. The two STIS observations (blue asterisks) match perfectly the ground-based line-intensity curve for the sampled phases in common ($\phi$\,$=$\,10.817 and $\phi$\,$=$\,12.839).

{ The ratio  [\ion{Fe}{iii}]\,$\lambda4659$\,$/$\,[\ion{Fe}{ii}]\,$\lambda4641$ was reported also for before the well-studied period 1990--2023 and is the main subject of this work. The dense monitoring in the 1992.5--2023 period is indicated by the brown line-intensity curve. It is similar to that involving the \ion{Fe}{ii}\,$\lambda4631$ line. The main difference is that the ground-based [\ion{Fe}{iii}]\,$\lambda4659$\,$/$\,[\ion{Fe}{ii}]\,$\lambda4641$ line-intensity ratio, is substantially higher in STIS (brown asterisks) than in ground-based (brown curve) data, although the two STIS observations have the same increasing slope as the ground-based observations for the specific orbital phases in common. The cycle-to-cycle evolution of the orbital maxima shows a gentle increase in contrast to the blue line-intensity ratio curve. It is shallower because [\ion{Fe}{ii}]\,$\lambda4641$ line intensity flux has been increasing almost at the same pace as [\ion{Fe}{iii}]\,$\lambda4659$ (see Figure \ref{Fig4}). This is not expected for a nebular regime in high density, as shown by \citet{Teodoro20} for the Weigelt C and D clumps. 

 We degraded (by convolving with a Gaussian profile) our spectra to a fixed R\,$=$\,5\,000 resolution to extend the temporal study to before the 1992.5--2023 period. This resolution corresponds to a degradation by a factor of $\sim$10$\times$ compared to the original resolution of most of the spectra (see Table~\ref{table_observat}). The red-filled circles on top of the brown line-ratio curve of Figure~\ref{Fig5} - (see Table~\ref{table_fe3fe2}) - are measurements in phases $\phi$\,$=$\,0.3\,$\pm$\,0.1. These are shown as red-filled circles connected with red dotted lines in Figure~\ref{Fig5}. The ratio values were $\sim$\,1.45 for the last three decades, slightly larger when compared to the historical values: $\sim$\,1.0 for the period 1960--90 and $<$\,1.0 in 1938--60. The ratio in 1974.08 and 1986.22 (orange empty triangles) was 1.05 and 1.15, respectively, slightly larger than those in 1961 and 1983 because they were taken at orbital phases closer to the maximum of the orbital modulation. 

Figure\,\ref{Fig5} shows a correspondence between the maxima in line-intensity ratio curves of [\ion{Fe}{iii}]\,$/$\,[\ion{Fe}{ii}] and [\ion{Fe}{iii}]\,$/$\,\ion{Fe}{ii} with the V$-$band light curve showing local photometric minima at $\phi$\,$\sim$\,0.8. The photometric modulation is getting more visible at later times. The cycle 9 was peculiar in both black and brown curves. The local minimum in the V$-$band was anomalously deep, corresponding to a high maximum in the Fe$^{++}$\,$/\,$ Fe$^{+}$ line-ratio curves.

\section{Discussion and Conclusions}
\label{sectiondiscussions}

We report long-term variability of three narrow iron lines, representative of intermediate and high-excitation in the ejecta clumps:  \ion{Fe}{ii}\,$\lambda4631$, [\ion{Fe}{ii}]\,$\lambda4641$, and [\ion{Fe}{iii}]\,$\lambda4659$. 

Absolute flux observed with \hst/STIS at the same phases, $\phi$\,$=$\,12.839 and $\phi$\,$=$\,10.817, separated by three cycles shows almost no variation for the Weigelt D clump, while the central star brightened by $\Delta\,$$_V$\,$\sim$\,1.27\ in the same period. This is in agreement with our hypothesis of a dissipating occulter in our LOS impacting the apparent brightness of the central star, with a much smaller or no impact on the Weigelt clumps.

Ground-based high-resolution spectra are reported for the last six orbital cycles, showing the long-term fading already reported by \citet{mehner15} and by \citet{damineli21, damineli23}. The line fluxes were accurately recovered after multiplying the line intensities by the continuum flux of the whole object (Homunculus \,$+$\, central star). The evolution of the normalized line fluxes is in fair agreement with coeval STIS absolute fluxes observations shown in Figure\,\ref{Fig4}. The \ion{Fe}{ii}\,$\lambda4631$ line flux remained constant since 2001. 
 [\ion{Fe}{ii}]\,$\lambda4641$ and [\ion{Fe}{iii}]\,$\lambda4659$ increased by $\sim$\,75\%. The ratio between [\ion{Fe}{iii}]\,$\lambda4659$ and [\ion{Fe}{ii}]\,$\lambda4641$ shows a small scatter Figure\,\ref{Fig5}). The effect of spectral resolution is demonstrated in Fig.~\ref{Fig5} by the shift from the brown curve corresponding to measurements in the original high-resolution spectra, as compared with the empty orange triangles measured after degrading the resolving power to R\,$=$\,5\,000, more similar to historical spectra. Our measurements connect well with those reported by \citet{humphreys08}, for phases $\phi$\,$=$\,0.3\,$\pm$\,0.1.
 
 The V$-$band lightcurve of the whole object (Homunculus\,$+$\,stellar core) shows a continuous brightening in the period 1938-2000, which has been interpreted as a clearing of dust close to the central system. The 1940s and 1990s "jumps" in the light curve have a clear counterpart in increasing ionization in Weigelt C and D knots. However, how could we explain that the ionization after 1990 did not increase at a similar pace as the stellar core brightening? This is possible if the material in the LOS of the Weigelt knots to the central star has its own dissipating timescale separate from the dissipating occulter between us and the central system. The UV narrow line absorption that disappeared just before 2018 \citep{gull23} indicates the existence of absorbing material that was also impacting our LOS to the central star, but not in between the Weigelt clumps and the central system.  

The long-term [\ion{Fe}{iii}]\,$\lambda4659$\,$/$\,[\ion{Fe}{ii}]\,$\lambda4641$ ratio increased by a substantial amount in cycle 1, then again around cycle  9. But in the last 4 cycles it increased by a very low rate - see Fig.\,\ref{Fig5}. The ionization degree increased slowly since 2001, because both [\ion{Fe}{ii}]\,$\lambda4641$ and [\ion{Fe}{iii}]\,$\lambda4659$ line fluxes increased at a similar rate -- $\sim$\,7\% per year -- since 2001 (see Fig.\,\ref{Fig4}).

The long-term increase of the [\ion{Fe}{ii}]\,$\lambda4641$ line was not expected if it was controlled only by collisional excitation in a high-density nebular regime, as discussed in \citet{Teodoro20}, which is $\log(N_e)$\,$\sim$\,7 for the Weigelt D knot. The fact that there have been temporary deeps, more noticeable in the last low excitation event, indicates some sensibility of this line to photoexcitation. This is not unexpected since the ground-based spectra cover an extended area, which could have lower-density regions (sensitive to photoexcitation) than the Weigelt D clump. Moreover, the clumps are also affected by extinction decrease in their LOS to the central stellar system.

For interpreting such a complex sequence of events, we present a schematic cartoon in  Fig.~\ref{cartoon}. An absorbing shell that disappeared in the 1940s (the dark red shell labeled by "4" would explain the simultaneous changes in the ionization of the Weigelt knots and the jump in the brightness of the entire Homunculus. The report that the brightness of the central star remained almost constant as the Homunculus jumped by $\sim$\,1\,mag \citep{thackeray53,oconnell56} is in accord with an omnidirectional extinction decrease. 

The increase in the 1990s excitation was different from that in the 1940s. The 1990s increasing excitation event occurred before the central core started to brighten in the 2000s, so the extinction decrease requires two components. Before 2000, the Homunculus was at least $\sim$\,10$\times$ brighter than the stellar core. After that, it remained at the same brightness \citep{damineli19}, the latter followed by the stellar brightening \citep{martin04, martin06, damineli19}. We represent the source of extinction in front of the Weigelt clumps that vanished as the gray cloud labeled "3" in Fig.\,\ref{cartoon}. It is possible that the dissipation of cloud "3" was part of a process of clearing the ejecta, which also impacted the material to our LOS cloud, which we call "the occulter" (labeled "1" and "2"). The occulter's dissipation core "1" may have started brightening in that event. The occulter halo "2", could also have started dissipating at that time, but we have no observations to check at what exact time before 2018 it finished its dissipation process \citep{gull23}.

\begin{figure}[ht]
\centering
{\includegraphics[width=1.05\linewidth, angle=0,viewport=0bp 10bp 920bp 745bp, clip]{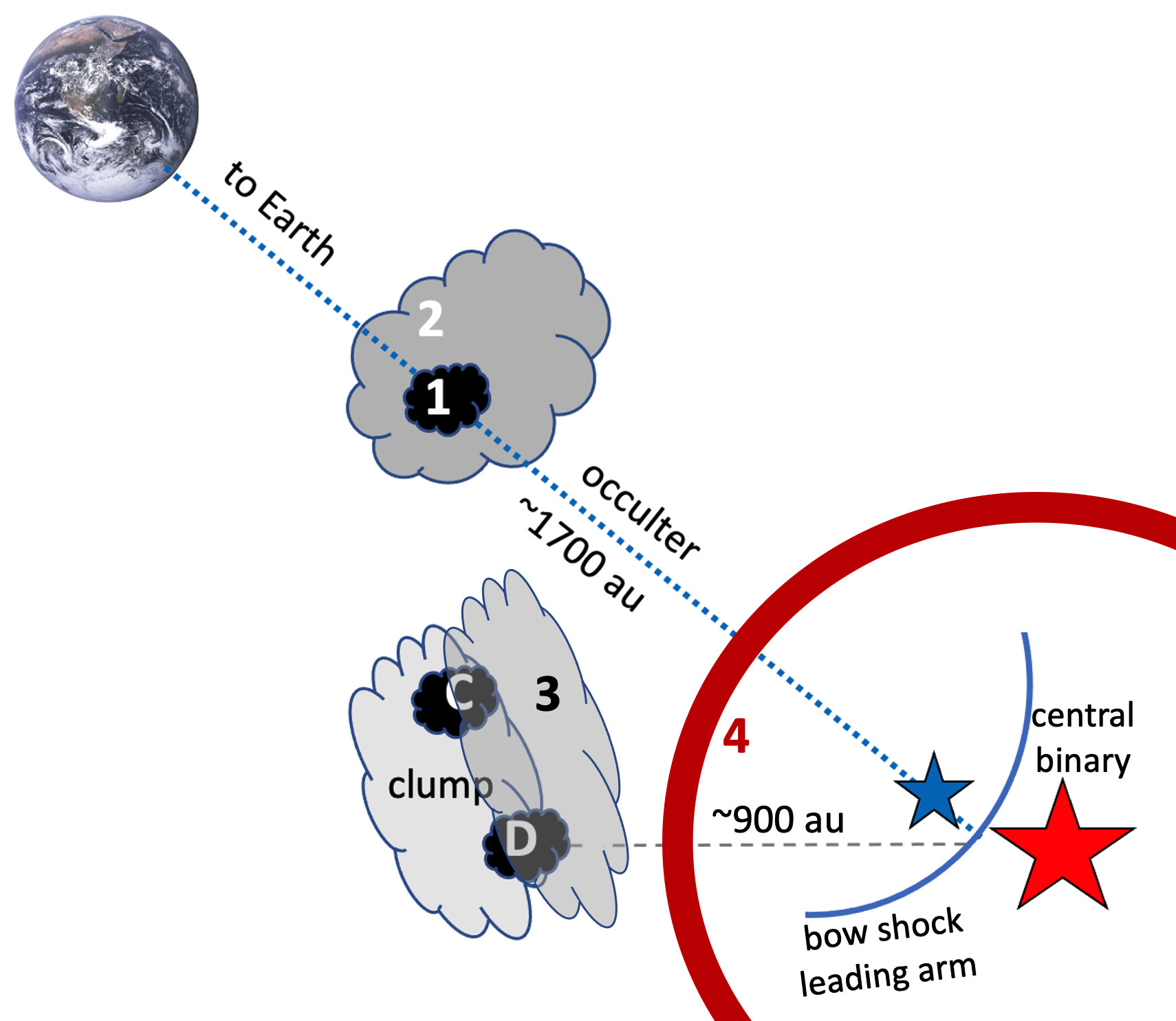}}
\caption{ Schematic cartoon showing the possible position of the extinction structures that have been dissipated. Shell "4" corresponds to the extinction structure that disappeared in the 1940s; clump "3" to the 1990s excitation jump. The cloud "2" represents the clump that was associated with the disappearance of narrow absorption lines around 1918 (occulter's halo). The knot "1" is the occulter core, which started dissipating around the 2000s.  Distances are not to scale, but the angle between our LOS to the central system and the LOS of the Weigelt clumps is 37 degrees \citep{Teodoro20}. }
\label{cartoon}
\end{figure}

\subsection{Extinction changes around the Weigelt clumps}
\label{subsection_extinction}

To deal with the extinction changes, we start by evaluating the extinction in our LOS to the Wegelt clumps by examining the flux increase in the [\ion{Fe}{ii}]\,$\lambda4641$ line, since this spectral line emission is in the nebular regime, as discussed in \citet{Teodoro20}, and is almost insensitive to changes due to illumination effects by the central system. This is well demonstrated by the steadiness of the [\ion{Fe}{ii}]\,$\lambda4815$ line flux { through the orbital cycle} - see their Fig.\,9.  In this approximation, the change in the flux of [\ion{Fe}{ii}]\,$\lambda4641$ line is due only to the extinction decrease to our LOS. The extinction in the time interval 1986.22-2002.5 should have decreased by $\Delta\,A_V$\,$\sim$\,0.3 (see Table\,\ref{table_extinction} in Appendix). This translates into an extinction decrease rate of $\sim$\,0.018\,mag.yr$^{-1}$. 
The [\ion{Fe}{iii}]\,$\lambda4659$ line flux suffers from the same extinction decrease so that the increase of ratio [\ion{Fe}{iii}]\,$\lambda4659$\,/\, [\ion{Fe}{ii}]\,$\lambda4641$ of $\sim$\,22\% in this period is due to increased exposure of the Weiglts clumps to the ionizing flux of the secondary star. This occurs if the extinction between the Weigelt clumps and the central binary has decreased. 

To evaluate the amount of the extinction decrease between the Weigelt clumps and the central system that caused the ionization ratio increase in 1986.22--2002.5 period we can use rough approximations. As before, we assume the high-density nebular regime in which the [\ion{Fe}{ii}]\,$\lambda4641$ line flux is constant and so the increase in the ionization ratio is totally due to the increase in the [\ion{Fe}{iii}]\,$\lambda4659$ line flux increase. Since this is a small increase, we assume that the increase in the ionizing flux incoming from the secondary star was of the same order. So, the decrease in extreme UV extinction is $\sim$\,0.21. If we adopt the reddening law used by \citet{hillier2001}, which is a combination of R\,$\sim$\,4.2 with a gray component, in the extreme UV (555\AA~) the extinction relations is A$_{UV}$\,$/$\,A$_V$\,$=$\,1.8. This implies in A$_V$\,$=$\,0.12. The exact amount of extinction decrease 
is difficult to assess, primarily because very close to the central system, dust grains are destroyed, and only very large grains remain for a longer time, requiring a combination of reddening laws. This changes with the distance to the star. A full model is out of the scope of this paper. 

For the time interval, 2002.5--2019.05 the extinction decrease of our LOS to Weigelt clumps was $\Delta\,A_V$\,$\sim$\,0.4, which led to an extinction decrease rate of $\sim$\,0.024\,mag.yr$^{-1}$. In this period, the [\ion{Fe}{iii}]\,$\lambda4659$\,$/$\,[\ion{Fe}{ii}]\,$\lambda4641$ ratio increased by just $\sim$\,10\%, which indicates a very small extinction decrease in the LOS of the Weigelt clumps to the central binary (see the flat behavior of this line ratio in Fig.\,\ref{Fig5}).

Just for comparison, using the magnitude brightening of the stellar core (following the method of \citealt{damineli19}), we found that the extinction caused by the "occulter" (clumps "1"\,$+$\,"2") in our LOS to the central system decreased by $\Delta\,A_V$\,$\sim$\,1.8 in the time interval 2002.5--2019.05 resulting in a rate of $\sim$\,0.11mag.yr$^{-1}$ to our LOS to the central star, which is about $\sim$\,4.6\,x faster than in our LOS to the Weigelt clumps.

 We further assume that these three recent events: a) the Weigelt clumps excitation increase in early 1990 (this work and \citealt{humphreys08}), b) the start of the occulter's dissipation around 2000 \citep{davidson99,martin06,damineli19}, and c) the disappearance of the UV narrow absorption lines just before 2018 \citep{gull23} are due to a common cause originated in the central system. Using the dates stated above and the distances in Figure\,\ref{cartoon} the perturbation started in 1980 and traveled at a speed of $\sim$\,400\,\kms. This is very close to the observed terminal wind speed of the primary star \citep{groh12}. Could an outburst in the primary star cause those mentioned above three recent events? We searched for a signal of wind variation using the H$\Delta\,$ line profile. Figure 3 of \citealt{damineli21} indicates no change since 1992.5 for phase-locked observations ($\phi$\,$=$\,0.02). We used one spectrum taken at the OPD observatory in 1998/May/13 ($\phi$\,$=$\,9.07) to compare with the one published by \citet{zanella1984} (1981/Dec/25, $\phi$\,$=$\,7.11). They show very similar line profiles, with no indication of a difference in the terminal speed. In addition, there is also no indication of exceptional photometric behavior in the whole object light curve in the period 1980--2000s. This indicates that the primary's wind has not changed appreciably throughout the last 40 years and the three events we are describing were not caused by a mass ejection of the primary star.

The fact that a) the three dusty clumps we described are located in the broad direction of the binary's major axis, which coincides with the wind-wind collision cavity orientation for most of the orbit; b) the cause of the changes in the excitation of the clumps traveled at the primary's wind terminal speed; c) the homunculus brightness over the last three decades remained constant; d) the global properties of the primary's wind have not changed over the last 40 years; suggesting that the described three events are "beamed", not involving a global change in the primary star.
We did not identify any mass ejection episode since 1980 that could have triggered these events. It looks as though it has been a silent cumulative work of the primary's wind blowing inside the wind-wind collision cavity and cleaning out the dust.

{
\section{Data Availability}
The complete version of Table~\ref{table_lineintensity} is available in the online journal in a machine-readable format following the CDS/Vizier standards.

\section{Acknowledgments}
AD thanks to CNPq (301490/2019-8) and FAPESP (2011/51680-6) for their support. NDR is grateful for support from the Cottrell Scholar Award \#CS-CSA-2023-143 sponsored by the Research Corporation for Science Advancement, as well as support from STScI's \hst GO programs 15611 and 15992. The work of FN is supported by NOIRLab, which is managed by the Association of Universities for Research in Astronomy (AURA) under a cooperative agreement with the National Science Foundation. AFJM is grateful for financial aid from NSERC (Canada).


{ Based on observations collected at the European Southern Observatory, Chile under Prog-IDs: UVES:  60.A-9022(A), 70.D-0607(A), 71.D-0168(A), 072.D-0524(A), 074.D-0141(A), 077.D-0618(A), 380.D-0036(A), 381.D-0004(A),282.D-5073(A, B, C, D, E), 089.D-0024(A), 592.D-0047(A, B, C). The first three programmes were described in \citet{stahl05} and the others in \citet{mehner15}; FEROS (partial list): 00.A-0000(A), 69.D-0378(A),, 69.D-0381(A), 71.D-0554(A),079.D-0564(C), 079.A-9201(A), 081.D-2008(A), 082.A-9208(A), 082.A-9209(A), 082.A-9210(A), 083.D-0589(A), 086.D-0997(A), 087.D-0946(A),089.D-0975(A), 098.A-9007(A). Spectra can be downloaded from  the ESO  database by using the instrument name (UVES or FEROS) plus the Julian Date provided in the Appendix Tables.}
Based on observations made at the Coud\'e focus of the 1.6-m telescope for the Observat\'orio do Pico dos Dias/LNA (Brazil).
Based in part on data from Mt. John University Observatory: MJUO -University of Canterbury - New Zealand).
Based in part on observations obtained through NOIRLab (formerly NOAO) allocations of NOAO-09B-153, NOAO-12A-216, NOAO-12B-194, NOAO-13B-328, NOAO-15A-0109, NOAO-18A-0295, NOAO-19B204, NOIRLab-20A-0054, and NOIRLab-21B-0334. This research has used data from the CTIO/SMARTS 1.5m telescope, which is operated as part of the SMARTS Consortium by RecONS (www.recons.org) members Todd Henry, Hodari James, Wei-Chun Jao, and Leonardo Paredes. At the telescope, observations were carried out by Roberto Aviles and Rodrigo.

\vspace{5mm}
\facilities{AAVSO, CTIO, LCOGT, ESO, MJUO, \hst, LNA, MSO}

\bibliographystyle{aasjournal}
\bibliography{refs} 
}

\appendix
\section{ EXTINCTION IN FRONT AND BEHIND CLUMPS "1"\,$+$\,"2", AND "3"}

To evaluate how much the extinction should have decreased in front of the clump "3" to our LOS, let us compare the evolution of the [\ion{Fe}{ii}]\,$\lambda4641$ line flux (see Table\,\ref{table_extinction}). We use the V$-$band magnitudes of the whole object (columns 2 and 3 or 3 and 5) - first row - to measure the magnitude difference in the stellar continuum between a pair of dates. Then we transform the magnitude difference to a flux ratio of the continuum and multiply it by the line intensity ratio (I\,$/$\,Ic) - second row - to derive the line intensity flux ratio (F\,$/$\,Fc). We calculate the magnitude difference corresponding to the flux ratio of the line for that pair of dates. We assume that there is no intrinsic evolution of the emission in the [\ion{Fe}{ii}]\,$\lambda4641$ line so the magnitude difference is the extinction difference - columns 4 and 6. The extinction increased by $\Delta\,A_V$\,$=$\,0.28\,mag in the time interval 1986.22--2002.5 and  $\Delta\,A_V$\,$=$\,0.40\,mag in the time interval 2002.5--2019.05. These values are for the extinction decrease in front of clump "3" to our LOS and are due to dissipation within the Homunculus.
 
For the stellar core, we can measure the extinction difference of the dissipating occulter "1"\,$+$\,"2") (see in Figure\,\ref{cartoon}) for the dates 2002.5--2019.05. We follow the procedure described by \citet{damineli19}, which is to start from the V$-$band magnitude of the whole object at one of the dates and subtract the flux of the Homunculus nebula, which is approximately constant (V$_H$\,$\sim$\,5.5) - \citet{davidson95} gives 5.7. We repeat the same calculation for the second date and translate the brightness increase into an extinction decrease to our LOS towards the central star. For the pair of dates 2002.5 and 2019.05 we obtain an extinction decrease $\Delta\,A_V$\,$=$\,1.8\,mag, which is $\sim$\,4.6\,x higher than that of the Weigelt clumps LOS to the central star in the same period. This extinction decrease adds the changes inside the clumps "1"\,$+$\,"2" to that within the Homunculus.

\begin{table}[!ht]
	\centering									
	\caption{Extinction decrease in our LOS to the Weigelt clumps and to the central star}
	\label{table_extinction}
	\begin{tabular}{llllll} 
    \hline							
	& 1986.22 &2002.5 &1986--2002 &2019.05& 2002--19 	\\
Parameter & Year & Year & $\Delta\,A_V$\,mag &Year& $\Delta\,A_V$\,mag \\
\hline
V$_{whole}$	 & 5.87 & 5.15 &- &4.31&-\\
I/Ic$_{[FeII]\,4641**}$& 2.4 & 2.14&0.28 & 1.43&0.4\\
V$_{core*}$ & - & 6.55 &-& 4.76  &1.8\\
\hline
\end{tabular}
\\
\textbf{Notes:}
(*) - Magnitudes of the core are obtained following \citet{damineli19} - see text.\\
(**) - After degrading spectral resolution to R\,$=$\,5\,000
\end{table}

It is possible to evaluate the total extinction of the occulter { (clumps "1"\,$+$\,"2" )} by using the extinction { in front of} the Weigelt BCD  knots reported by \citet{davidson95} - $A_V$\,$\sim$\,2.0 - using data taken with \hst in 1991.4. By subtracting the { total} extinction decrease from that date to 2019.5, the total extinction { (interstellar plus within the Homunmculus)} in our LOS to the Weigelt clumps should have been $A_V$\,$\sim$\,1.44. For the measured stellar core brightness, V\,$=$\,4.76 - see Table\, \ref{table_extinction} - the brightness after correcting for the same material that affects the Weigelt clumps (Homunculus plus interstellar) would be V\,$=$\,3.32. Using the unreddened apparent magnitude from \citet{hillier2001} model, V$_0$\,$=$\,0.94, the extinction caused by the occulter to our LOS should have been $A_V$\,$\sim$\,2.38 in 2019.5. The extinction decrease since 1991.4, reported by \citet{davidson95} - $A_V$\,$\sim$\,6.1 - should have been { $\Delta\,A_V$\,$\sim$\,3.88} which translates into an average rate of 0.14\,mag.yr$^{-1}$, in good agreement with the brightening rate (0.13\,mag.yr$^{-1}$) reported by \citet{damineli19} for the period 1998.2--2015.7. At this rate, the dissipation of the occulter { (no more extinction decrease)} would be completed in 2040.

\section{Extended data tables (online material)}
\label{tab_lineintensity}

\startlongtable
\begin{deluxetable}{ccccccccc}
\tablecaption{Normalized line intensities.}	
\label{table_lineintensity}
\tablenum{1}
\tablehead{\colhead{row} & \colhead{Observat } & \colhead{JD} & \colhead{year } & \colhead{phase } & \colhead{FeII } & \colhead{[FeII]} & \colhead{[FeIII]} & \colhead{V$_{flux}$} \\ 
\colhead{} & \colhead{} & \colhead{} & \colhead{} & \colhead{(14$=$2020.2)} & \colhead{(4631$\AA$)} & \colhead{(4641$\AA$)} & \colhead{(4659$\AA$)} & \colhead{(norm)} }
\startdata
1	&	ESO/FEROS	&	2448793.99	&	1992.4675	&	9.00514	&	3.14	&	2.79	&	1.33	&	0.71	\\
2	&	ESO/FEROS	&	2448825.01	&	1992.5524	&	9.02047	&	3.01	&	2.48	&	1.15	&	0.71	\\
3	&	ESO/FEROS	&	2448830.02	&	1992.5661	&	9.02295	&	3.21	&	2.55	&	1.27	&	0.71	\\
4	&	ESO/FEROS	&	2448839.00	&	1992.5907	&	9.02739	&	3.35	&	2.52	&	1.25	&	0.71	\\
5	&	ESO/FEROS	&	2448843.99	&	1992.6043	&	9.02986	&	3.23	&	2.32	&	1.18	&	0.71	\\
6	&	ESO/FEROS	&	2449063.11	&	1993.2043	&	9.13819	&	7.05	&	3.52	&	3.10	&	0.65	\\
7	&	ESO/FEROS	&	2449068.11	&	1993.2180	&	9.14066	&	7.06	&	3.58	&	3.11	&	0.65	\\
8	&	ESO/FEROS	&	2449073.15	&	1993.2318	&	9.14315	&	7.16	&	3.43	&	3.09	&	0.65	\\
9	&	ESO/FEROS	&	2449078.11	&	1993.2453	&	9.14560	&	6.44	&	3.22	&	3.09	&	0.65	\\
10	&	ESO/FEROS	&	2449083.10	&	1993.2590	&	9.14807	&	6.34	&	3.19	&	2.96	&	0.65	\\
11	&	ESO/FEROS	&	2449088.11	&	1993.2727	&	9.15055	&	6.21	&	3.00	&	2.96	&	0.65	\\
12	&	ESO/FEROS	&	2449095.09	&	1993.2918	&	9.15400	&	6.40	&	3.00	&	2.90	&	0.65	\\
13	&	ESO/FEROS	&	2449102.11	&	1993.3110	&	9.15747	&	6.81	&	3.53	&	3.32	&	0.65	\\
14	&	ESO/FEROS	&	2449106.11	&	1993.3220	&	9.15944	&	6.94	&	3.47	&	3.36	&	0.65	\\
15	&	ESO/FEROS	&	2449110.18	&	1993.3332	&	9.16146	&	6.16	&	3.24	&	3.07	&	0.65	\\
16	&	ESO/FEROS	&	2449113.08	&	1993.3411	&	9.16289	&	5.91	&	3.39	&	3.00	&	0.65	\\
17	&	ESO/FEROS	&	2449114.18	&	1993.3441	&	9.16343	&	5.62	&	3.14	&	5.62	&	0.65	\\
18	&	ESO/FEROS	&	2449128.04	&	1993.3820	&	9.17029	&	6.07	&	3.07	&	3.09	&	0.65	\\
19	&	ESO/FEROS	&	2449138.09	&	1993.4096	&	9.17526	&	6.20	&	3.24	&	3.13	&	0.65	\\
20	&	ESO/FEROS	&	2449395.37	&	1994.1139	&	9.30245	&	6.63	&	3.41	&	3.43	&	0.66	\\
21	&	ESO/FEROS	&	2449408.03	&	1994.1486	&	9.30871	&	5.00	&	2.92	&	3.50	&	0.66	\\
22	&	ESO/FEROS	&	2449424.02	&	1994.1924	&	9.31662	&	4.94	&	2.62	&	3.45	&	0.66	\\
23	&	ESO/FEROS	&	2449448.10	&	1994.2583	&	9.32852	&	4.93	&	2.62	&	3.58	&	0.66	\\
24	&	ESO/FEROS	&	2449467.06	&	1994.3102	&	9.33790	&	5.00	&	2.69	&	3.50	&	0.66	\\
25	&	ESO/FEROS	&	2449472.10	&	1994.3240	&	9.34039	&	4.44	&	2.54	&	3.32	&	0.66	\\
26	&	ESO/FEROS	&	2449514.04	&	1994.4389	&	9.36112	&	4.77	&	2.67	&	3.47	&	0.66	\\
27	&	ESO/FEROS	&	2449520.11	&	1994.4555	&	9.36412	&	5.16	&	2.81	&	3.73	&	0.66	\\
28	&	ESO/FEROS	&	2449522.04	&	1994.4607	&	9.36508	&	5.22	&	2.79	&	3.69	&	0.66	\\
29	&	ESO/FEROS	&	2449525.03	&	1994.4689	&	9.36655	&	5.55	&	2.97	&	4.03	&	0.65	\\
30	&	ESO/FEROS	&	2449527.03	&	1994.4744	&	9.36754	&	5.15	&	2.82	&	3.71	&	0.65	\\
31	&	ESO/FEROS	&	2449532.02	&	1994.4881	&	9.37001	&	4.65	&	2.59	&	3.56	&	0.65	\\
32	&	ESO/FEROS	&	2449770.24	&	1995.1403	&	9.48778	&	5.17	&	2.70	&	4.78	&	0.58	\\
33	&	ESO/FEROS	&	2449786.12	&	1995.1838	&	9.49563	&	5.52	&	2.79	&	4.99	&	0.58	\\
34	&	ESO/FEROS	&	2449800.24	&	1995.2224	&	9.50262	&	5.02	&	2.68	&	4.63	&	0.58	\\
35	&	ESO/FEROS	&	2449836.05	&	1995.3205	&	9.52032	&	5.51	&	2.88	&	4.94	&	0.58	\\
36	&	ESO/FEROS	&	2449846.04	&	1995.3478	&	9.52526	&	4.87	&	2.52	&	4.60	&	0.58	\\
37	&	ESO/FEROS	&	2449860.07	&	1995.3862	&	9.53219	&	5.15	&	2.81	&	3.82	&	0.57	\\
38	&	ESO/FEROS	&	2450120.28	&	1996.0987	&	9.66084	&	5.73	&	3.03	&	5.78	&	0.59	\\
39	&	ESO/FEROS	&	2450134.27	&	1996.1370	&	9.66776	&	5.60	&	3.05	&	5.75	&	0.59	\\
40	&	ESO/FEROS	&	2450146.37	&	1996.1701	&	9.67374	&	5.60	&	2.99	&	5.63	&	0.59	\\
41	&	ESO/FEROS	&	2450160.25	&	1996.2081	&	9.68060	&	5.70	&	3.05	&	5.70	&	0.59	\\
42	&	ESO/FEROS	&	2450174.26	&	1996.2464	&	9.68753	&	5.89	&	3.25	&	5.80	&	0.59	\\
43	&	ESO/FEROS	&	2450188.31	&	1996.2849	&	9.69447	&	6.26	&	3.28	&	5.80	&	0.59	\\
44	&	ESO/FEROS	&	2450201.23	&	1996.3203	&	9.70086	&	5.05	&	3.14	&	5.75	&	0.59	\\
45	&	LNA/OPD	&	2450211.50	&	1996.3484	&	9.70594	&	1.53	&	1.20	&	1.81	&	0.59	\\
46	&	LNA/OPD	&	2450211.50	&	1996.3484	&	9.70594	&	1.70	&	1.26	&	1.84	&	0.59	\\
47	&	ESO/FEROS	&	2450237.97	&	1996.4209	&	9.71902	&	5.00	&	3.01	&	5.70	&	0.59	\\
48	&	ESO/FEROS	&	2450238.01	&	1996.4210	&	9.71905	&	5.10	&	3.00	&	5.78	&	0.59	\\
49	&	ESO/FEROS	&	2450241.95	&	1996.4318	&	9.72099	&	5.20	&	3.05	&	5.80	&	0.59	\\
50	&	ESO/FEROS	&	2450242.02	&	1996.4319	&	9.72103	&	5.50	&	2.94	&	5.72	&	0.59	\\
51	&	ESO/FEROS	&	2450244.08	&	1996.4376	&	9.72205	&	5.14	&	2.84	&	5.25	&	0.59	\\
52	&	ESO/FEROS	&	2450449.25	&	1996.9993	&	9.82348	&	5.29	&	2.75	&	5.11	&	0.62	\\
53	&	ESO/FEROS	&	2450450.26	&	1997.0021	&	9.82398	&	5.10	&	2.90	&	5.40	&	0.62	\\
54	&	ESO/FEROS	&	2450451.21	&	1997.0047	&	9.82445	&	5.36	&	3.33	&	4.90	&	0.62	\\
55	&	ESO/FEROS	&	2450452.21	&	1997.0074	&	9.82494	&	6.08	&	3.30	&	4.85	&	0.62	\\
56	&	ESO/FEROS	&	2450453.19	&	1997.0101	&	9.82543	&	6.79	&	3.37	&	4.88	&	0.62	\\
57	&	ESO/FEROS	&	2450454.22	&	1997.0129	&	9.82594	&	5.97	&	3.28	&	4.76	&	0.62	\\
58	&	ESO/FEROS	&	2450455.30	&	1997.0159	&	9.82647	&	6.49	&	3.34	&	4.90	&	0.62	\\
59	&	ESO/FEROS	&	2450456.22	&	1997.0184	&	9.82692	&	6.44	&	3.30	&	4.90	&	0.62	\\
60	&	ESO/FEROS	&	2450457.18	&	1997.0210	&	9.82740	&	6.11	&	3.43	&	4.79	&	0.62	\\
61	&	ESO/FEROS	&	2450464.34	&	1997.0406	&	9.83094	&	6.14	&	3.32	&	4.83	&	0.62	\\
62	&	ESO/FEROS	&	2450472.35	&	1997.0626	&	9.83490	&	6.20	&	3.28	&	4.96	&	0.62	\\
63	&	ESO/FEROS	&	2450476.33	&	1997.0734	&	9.83686	&	7.00	&	3.42	&	5.00	&	0.62	\\
64	&	ESO/FEROS	&	2450516.18	&	1997.1826	&	9.85657	&	7.10	&	3.41	&	5.00	&	0.62	\\
65	&	ESO/FEROS	&	2450518.13	&	1997.1879	&	9.85753	&	6.67	&	3.48	&	5.08	&	0.62	\\
66	&	ESO/FEROS	&	2450520.13	&	1997.1934	&	9.85852	&	5.34	&	3.33	&	4.20	&	0.62	\\
67	&	ESO/FEROS	&	2450522.19	&	1997.1990	&	9.85954	&	5.30	&	2.98	&	3.95	&	0.62	\\
68	&	ESO/FEROS	&	2450524.20	&	1997.2045	&	9.86053	&	5.75	&	3.13	&	4.18	&	0.62	\\
69	&	ESO/FEROS	&	2450527.12	&	1997.2125	&	9.86198	&	5.70	&	3.18	&	4.18	&	0.62	\\
70	&	ESO/FEROS	&	2450531.14	&	1997.2235	&	9.86396	&	5.65	&	3.13	&	4.20	&	0.62	\\
71	&	ESO/FEROS	&	2450533.11	&	1997.2289	&	9.86494	&	5.73	&	3.16	&	4.15	&	0.62	\\
72	&	ESO/FEROS	&	2450535.17	&	1997.2345	&	9.86596	&	5.56	&	3.26	&	4.15	&	0.62	\\
73	&	ESO/FEROS	&	2450537.15	&	1997.2400	&	9.86694	&	5.76	&	3.08	&	4.17	&	0.62	\\
74	&	ESO/FEROS	&	2450539.15	&	1997.2454	&	9.86792	&	5.68	&	3.16	&	4.15	&	0.62	\\
75	&	ESO/FEROS	&	2450541.16	&	1997.2510	&	9.86892	&	5.81	&	3.19	&	4.13	&	0.62	\\
76	&	ESO/FEROS	&	2450543.16	&	1997.2564	&	9.86991	&	5.90	&	3.27	&	4.25	&	0.62	\\
77	&	ESO/FEROS	&	2450545.14	&	1997.2619	&	9.87089	&	5.82	&	3.19	&	4.22	&	0.62	\\
78	&	ESO/FEROS	&	2450547.15	&	1997.2673	&	9.87188	&	5.94	&	3.27	&	4.21	&	0.62	\\
79	&	ESO/FEROS	&	2450549.15	&	1997.2728	&	9.87287	&	5.76	&	3.25	&	4.21	&	0.62	\\
80	&	ESO/FEROS	&	2450551.16	&	1997.2783	&	9.87386	&	5.70	&	3.20	&	4.17	&	0.62	\\
81	&	ESO/FEROS	&	2450553.16	&	1997.2838	&	9.87485	&	6.00	&	3.40	&	4.45	&	0.62	\\
82	&	ESO/FEROS	&	2450556.08	&	1997.2918	&	9.87629	&	5.93	&	3.38	&	4.35	&	0.62	\\
83	&	ESO/FEROS	&	2450557.11	&	1997.2946	&	9.87680	&	5.88	&	3.13	&	4.22	&	0.62	\\
84	&	ESO/FEROS	&	2450559.08	&	1997.3000	&	9.87778	&	6.39	&	3.39	&	4.51	&	0.62	\\
85	&	ESO/FEROS	&	2450561.08	&	1997.3055	&	9.87876	&	7.28	&	3.64	&	4.80	&	0.62	\\
86	&	ESO/FEROS	&	2450563.09	&	1997.3110	&	9.87976	&	6.92	&	3.54	&	4.72	&	0.62	\\
87	&	ESO/FEROS	&	2450565.08	&	1997.3164	&	9.88074	&	8.10	&	3.84	&	5.20	&	0.62	\\
88	&	LNA/OPD	&	2450652.46	&	1997.5557	&	9.92394	&	1.58	&	1.20	&	1.83	&	0.67	\\
89	&	LNA/OPD	&	2450652.50	&	1997.5558	&	9.92396	&	1.56	&	1.21	&	1.87	&	0.67	\\
90	&	LNA/OPD	&	2450761.833	&	1997.8551	&	9.97802	&	2.00	&	1.41	&	1.95	&	0.67	\\
91	&	LNA/OPD	&	2450762.50	&	1997.8569	&	9.97835	&	1.61	&	1.20	&	1.79	&	0.67	\\
92	&	LNA/OPD	&	2450762.79	&	1997.8577	&	9.97849	&	1.57	&	1.21	&	1.87	&	0.67	\\
93	&	LNA/OPD	&	2450762.79	&	1997.8577	&	9.97849	&	1.58	&	1.20	&	1.76	&	0.67	\\
94	&	MSO	&	2450771.20	&	1997.8808	&	9.98265	&	1.39	&	1.33	&	1.02	&	0.67	\\
95	&	MSO	&	2450797.14	&	1997.9518	&	9.99547	&	1.40	&	1.31	&	1.02	&	0.77	\\
96	&	MSO	&	2450808.12	&	1997.9818	&	10.00090	&	1.36	&	1.29	&	1.01	&	0.77	\\
97	&	MSO	&	2450870.10	&	1998.1515	&	10.03154	&	1.39	&	1.32	&	1.01	&	0.77	\\
98	&	MSO	&	2450884.00	&	1998.1896	&	10.03841	&	1.34	&	1.29	&	1.01	&	0.77	\\
99	&	MSO	&	2450896.04	&	1998.2226	&	10.04437	&	1.42	&	1.32	&	1.02	&	0.77	\\
100	&	LNA/OPD	&	2450946.46	&	1998.3606	&	10.06929	&	1.61	&	1.21	&	1.84	&	0.77	\\
101	&	LNA/OPD	&	2450946.50	&	1998.3607	&	10.06931	&	1.47	&	1.20	&	1.72	&	0.71	\\
102	&	ESO/FEROS	&	2451173.23	&	1998.9815	&	10.18140	&	7.64	&	3.77	&	5.13	&	0.84	\\
103	&	MSO	&	2451188.10	&	1999.0222	&	10.18876	&	1.55	&	1.40	&	1.10	&	0.84	\\
104	&	ESO/FEROS	&	2451381.98	&	1999.5530	&	10.28461	&	7.38	&	3.54	&	4.70	&	0.84	\\
105	&	ESO/FEROS	&	2451391.97	&	1999.5803	&	10.28955	&	4.57	&	2.55	&	3.10	&	0.98	\\
106	&	ESO/FEROS	&	2451594.41	&	2000.1346	&	10.38963	&	1.97	&	1.48	&	1.43	&	0.98	\\
107	&	ESO/FEROS	&	2451654.05	&	2000.2979	&	10.41912	&	2.00	&	1.57	&	1.49	&	0.98	\\
108	&	ESO/FEROS	&	2452010.97	&	2001.2751	&	10.59557	&	2.09	&	1.59	&	1.40	&	0.98	\\
109	&	ESO/FEROS	&	2452301.19	&	2002.0697	&	10.73906	&	2.15	&	1.57	&	1.32	&	1.00	\\
110	&	LNA/OPD	&	2452460.375	&	2002.5055	&	10.81776	&	2.50	&	1.54	&	2.05	&	1.00	\\
111	&	ESO/UVES	&	2452615.32	&	2002.9297	&	10.89436	&	1.59	&	1.23	&	1.88	&	1.00	\\
112	&	ESO/UVES	&	2452620.34	&	2002.9434	&	10.89684	&	1.49	&	1.15	&	1.77	&	1.00	\\
113	&	ESO/UVES	&	2452684.13	&	2003.1181	&	10.92838	&	1.52	&	1.15	&	1.74	&	0.96	\\
114	&	LNA/OPD	&	2452773.54	&	2003.3629	&	10.97258	&	1.85	&	1.37	&	1.96	&	0.96	\\
115	&	ESO/UVES	&	2452793.98	&	2003.4188	&	10.98269	&	1.59	&	1.20	&	1.87	&	0.96	\\
116	&	LNA/OPD	&	2452810.52	&	2003.4641	&	10.99086	&	1.95	&	1.45	&	1.33	&	0.96	\\
117	&	LNA/OPD	&	2452811.52	&	2003.4669	&	10.99136	&	1.95	&	1.49	&	1.36	&	0.96	\\
118	&	LNA/OPD	&	2452812.52	&	2003.4696	&	10.99185	&	1.96	&	1.44	&	1.48	&	0.96	\\
119	&	LNA/OPD	&	2452813.52	&	2003.4723	&	10.99235	&	1.94	&	1.44	&	1.52	&	0.96	\\
120	&	LNA/OPD	&	2452814.52	&	2003.4751	&	10.99284	&	1.95	&	1.44	&	1.67	&	0.96	\\
121	&	LNA/OPD	&	2452814.52	&	2003.4751	&	10.99284	&	1.83	&	1.34	&	1.70	&	0.96	\\
122	&	LNA/OPD	&	2452815.52	&	2003.4778	&	10.99334	&	1.87	&	1.35	&	1.72	&	0.96	\\
123	&	LNA/OPD	&	2452815.52	&	2003.4778	&	10.99334	&	2.02	&	1.45	&	2.07	&	0.96	\\
124	&	LNA/OPD	&	2452816.52	&	2003.4805	&	10.99383	&	2.02	&	1.46	&	2.07	&	0.96	\\
125	&	LNA/OPD	&	2452817.52	&	2003.4833	&	10.99432	&	1.97	&	1.47	&	2.07	&	0.96	\\
126	&	LNA/OPD	&	2452819.52	&	2003.4888	&	10.99531	&	1.94	&	1.42	&	2.04	&	0.96	\\
127	&	LNA/OPD	&	2452820.52	&	2003.4915	&	10.99581	&	1.92	&	1.37	&	2.01	&	0.96	\\
128	&	LNA/OPD	&	2452822.06	&	2003.4957	&	10.99657	&	1.44	&	1.33	&	1.02	&	0.96	\\
129	&	LNA/OPD	&	2452822.38	&	2003.4966	&	10.99672	&	2.14	&	1.51	&	1.85	&	0.96	\\
130	&	LNA/OPD	&	2452822.41	&	2003.4967	&	10.99674	&	1.48	&	1.32	&	1.02	&	0.96	\\
131	&	LNA/OPD	&	2452823.375	&	2003.4993	&	10.99722	&	2.29	&	1.57	&	1.90	&	0.96	\\
132	&	LNA/OPD	&	2452823.77	&	2003.5004	&	10.99741	&	1.45	&	1.27	&	1.01	&	0.96	\\
133	&	LNA/OPD	&	2452823.77	&	2003.5004	&	10.99741	&	1.58	&	1.36	&	1.00	&	0.96	\\
134	&	LNA/OPD	&	2452824.375	&	2003.5021	&	10.99771	&	2.45	&	1.60	&	2.13	&	0.96	\\
135	&	ESO/UVES	&	2452825.01	&	2003.5038	&	10.99803	&	1.58	&	1.23	&	1.90	&	0.96	\\
136	&	LNA/OPD	&	2452825.97	&	2003.5064	&	10.99850	&	1.69	&	1.31	&	1.00	&	0.96	\\
137	&	LNA/OPD	&	2452826.375	&	2003.5075	&	10.99870	&	2.22	&	1.51	&	2.03	&	0.96	\\
138	&	LNA/OPD	&	2452826.375	&	2003.5075	&	10.99870	&	2.20	&	1.61	&	2.10	&	0.96	\\
139	&	LNA/OPD	&	2452827.375	&	2003.5103	&	10.99920	&	2.16	&	1.55	&	1.64	&	0.96	\\
140	&	LNA/OPD	&	2452827.98	&	2003.5119	&	10.99950	&	1.58	&	1.30	&	0.98	&	0.96	\\
141	&	LNA/OPD	&	2452828.38	&	2003.5130	&	10.99969	&	2.06	&	1.51	&	1.56	&	0.96	\\
142	&	LNA/OPD	&	2452829.98	&	2003.5174	&	11.00048	&	1.57	&	1.28	&	0.98	&	0.96	\\
143	&	LNA/OPD	&	2452830.38	&	2003.5185	&	11.00068	&	1.97	&	1.43	&	1.42	&	1.15	\\
144	&	LNA/OPD	&	2452840.98	&	2003.5475	&	11.00592	&	1.58	&	1.29	&	1.01	&	1.15	\\
145	&	LNA/OPD	&	2452987.71	&	2003.9492	&	11.07846	&	1.45	&	1.31	&	1.02	&	1.15	\\
146	&	LNA/OPD	&	2452987.833	&	2003.9496	&	11.07853	&	2.40	&	1.70	&	2.06	&	1.15	\\
147	&	ESO/UVES	&	2453413.39	&	2005.1147	&	11.28891	&	1.50	&	1.15	&	1.80	&	1.25	\\
148	&	ESO/UVES	&	2453834.11	&	2006.2666	&	11.49692	&	1.47	&	1.18	&	1.82	&	1.51	\\
149	&	ESO/UVES	&	2454475.27	&	2008.0220	&	11.81390	&	1.45	&	1.17	&	1.75	&	1.17	\\
150	&	ESO/UVES	&	2454513.30	&	2008.1261	&	11.83270	&	1.59	&	1.24	&	1.91	&	1.17	\\
151	&	ESO/UVES	&	2454535.28	&	2008.1863	&	11.84357	&	1.54	&	1.21	&	1.80	&	1.17	\\
152	&	ESO/UVES	&	2454554.28	&	2008.2383	&	11.85296	&	1.61	&	1.20	&	1.97	&	1.17	\\
153	&	ESO/UVES	&	2454567.01	&	2008.2731	&	11.85925	&	1.53	&	1.24	&	1.90	&	1.17	\\
154	&	ESO/UVES	&	2454583.02	&	2008.3170	&	11.86717	&	1.67	&	1.27	&	2.00	&	1.17	\\
155	&	LNA/OPD	&	2454596.417	&	2008.3536	&	11.87379	&	2.41	&	1.48	&	2.05	&	1.17	\\
156	&	LNA/OPD	&	2454596.50	&	2008.3539	&	11.87383	&	1.90	&	1.45	&	2.07	&	1.17	\\
157	&	LNA/OPD	&	2454598.40	&	2008.3591	&	11.87477	&	1.83	&	1.37	&	1.97	&	1.17	\\
158	&	ESO/FEROS	&	2454601.02	&	2008.3663	&	11.87607	&	1.85	&	1.44	&	1.14	&	1.17	\\
159	&	ESO/UVES	&	2454629.02	&	2008.4429	&	11.88991	&	1.70	&	1.22	&	2.05	&	1.17	\\
160	&	ESO/UVES	&	2454656.03	&	2008.5169	&	11.90326	&	1.68	&	1.31	&	2.14	&	1.19	\\
161	&	ESO/FEROS	&	2454821.33	&	2008.9694	&	11.98499	&	1.78	&	1.42	&	1.09	&	1.36	\\
162	&	ESO/FEROS	&	2454822.37	&	2008.9723	&	11.98550	&	1.90	&	1.45	&	1.37	&	1.36	\\
163	&	ESO/FEROS	&	2454823.29	&	2008.9748	&	11.98596	&	1.64	&	1.37	&	1.05	&	1.36	\\
164	&	ESO/FEROS	&	2454824.31	&	2008.9776	&	11.98646	&	1.72	&	1.42	&	1.05	&	1.36	\\
165	&	ESO/FEROS	&	2454825.30	&	2008.9803	&	11.98695	&	1.62	&	1.39	&	1.05	&	1.36	\\
166	&	ESO/FEROS	&	2454827.37	&	2008.9859	&	11.98797	&	1.59	&	1.33	&	1.03	&	1.36	\\
167	&	ESO/FEROS	&	2454828.36	&	2008.9887	&	11.98846	&	1.45	&	1.35	&	1.02	&	1.36	\\
168	&	ESO/FEROS	&	2454829.37	&	2008.9914	&	11.98896	&	1.80	&	1.40	&	1.11	&	1.36	\\
169	&	ESO/FEROS	&	2454830.37	&	2008.9942	&	11.98946	&	1.65	&	1.47	&	1.04	&	1.36	\\
170	&	ESO/FEROS	&	2454831.37	&	2008.9969	&	11.98995	&	1.56	&	1.37	&	1.02	&	1.36	\\
171	&	ESO/FEROS	&	2454832.37	&	2008.9996	&	11.99044	&	1.60	&	1.44	&	1.08	&	1.36	\\
172	&	ESO/FEROS	&	2454833.37	&	2009.0024	&	11.99094	&	1.52	&	1.36	&	1.02	&	1.36	\\
173	&	ESO/FEROS	&	2454834.38	&	2009.0051	&	11.99144	&	1.55	&	1.35	&	1.48	&	1.36	\\
174	&	ESO/Hexapod	&	2454844.36	&	2009.0325	&	11.99637	&	2.24	&	1.49	&	2.14	&	1.36	\\
175	&	ESO/Hexapod	&	2454845.36	&	2009.0352	&	11.99686	&	2.02	&	1.42	&	1.96	&	1.36	\\
176	&	LNA/OPD	&	2454882.63	&	2009.1372	&	12.01529	&	1.85	&	1.39	&	2.07	&	1.36	\\
177	&	ESO/Hexapod	&	2454897.50	&	2009.1780	&	12.02264	&	2.33	&	1.57	&	2.13	&	1.36	\\
178	&	LNA/OPD	&	2454940.48	&	2009.2956	&	12.04389	&	1.87	&	1.39	&	2.06	&	1.36	\\
179	&	ESO/FEROS	&	2454953.10	&	2009.3302	&	12.05013	&	1.48	&	1.31	&	1.01	&	1.36	\\
180	&	ESO/FEROS	&	2454955.08	&	2009.3356	&	12.05111	&	1.40	&	1.31	&	1.02	&	1.44	\\
181	&	CTIO/CHIRON	&	2455989.69	&	2012.1682	&	12.56261	&	2.49	&	1.58	&	2.37	&	1.44	\\
182	&	CTIO/CHIRON	&	2455993.75	&	2012.1793	&	12.56462	&	2.54	&	1.63	&	2.45	&	1.44	\\
183	&	CTIO/CHIRON	&	2456001.73	&	2012.2012	&	12.56856	&	2.47	&	1.57	&	2.33	&	1.44	\\
184	&	CTIO/CHIRON	&	2456013.69	&	2012.2339	&	12.57447	&	2.65	&	1.65	&	2.52	&	1.44	\\
185	&	CTIO/CHIRON	&	2456221.89	&	2012.8039	&	12.67741	&	2.48	&	1.56	&	2.30	&	1.44	\\
186	&	CTIO/CHIRON	&	2456236.84	&	2012.8449	&	12.68480	&	2.70	&	1.55	&	2.40	&	1.44	\\
187	&	CTIO/CHIRON	&	2456238.87	&	2012.8504	&	12.68580	&	2.71	&	1.66	&	2.49	&	1.44	\\
188	&	CTIO/CHIRON	&	2456248.81	&	2012.8777	&	12.69072	&	2.55	&	1.63	&	2.32	&	1.44	\\
189	&	CTIO/CHIRON	&	2456254.88	&	2012.8943	&	12.69372	&	2.80	&	1.64	&	2.50	&	1.44	\\
190	&	CTIO/CHIRON	&	2456260.78	&	2012.9104	&	12.69663	&	2.80	&	1.64	&	2.50	&	1.73	\\
191	&	CTIO/CHIRON	&	2456275.82	&	2012.9516	&	12.70407	&	2.70	&	1.64	&	2.44	&	1.73	\\
192	&	CTIO/CHIRON	&	2456289.80	&	2012.9899	&	12.71098	&	2.82	&	1.63	&	2.50	&	1.73	\\
193	&	CTIO/CHIRON	&	2456361.65	&	2013.1866	&	12.74650	&	2.67	&	1.63	&	2.39	&	1.73	\\
194	&	CTIO/CHIRON	&	2456401.59	&	2013.2959	&	12.76625	&	2.82	&	1.77	&	2.67	&	1.73	\\
195	&	CTIO/CHIRON	&	2456417.59	&	2013.3397	&	12.77416	&	2.65	&	1.63	&	2.31	&	1.62	\\
196	&	CTIO/CHIRON	&	2456607.84	&	2013.8606	&	12.86822	&	2.86	&	1.58	&	2.40	&	1.58	\\
197	&	CTIO/CHIRON	&	2456612.87	&	2013.8744	&	12.87070	&	2.80	&	1.66	&	2.52	&	1.58	\\
198	&	CTIO/CHIRON	&	2456656.70	&	2013.9944	&	12.89237	&	2.80	&	1.66	&	2.48	&	1.64	\\
199	&	CTIO/CHIRON	&	2456659.71	&	2014.0026	&	12.89386	&	2.81	&	1.71	&	2.56	&	1.64	\\
200	&	CTIO/CHIRON	&	2456664.67	&	2014.0162	&	12.89631	&	2.83	&	1.79	&	2.66	&	1.64	\\
201	&	CTIO/CHIRON	&	2456670.78	&	2014.0329	&	12.89933	&	2.78	&	1.69	&	2.35	&	1.64	\\
202	&	CTIO/CHIRON	&	2456672.84	&	2014.0386	&	12.90035	&	2.82	&	1.77	&	2.60	&	1.64	\\
203	&	CTIO/CHIRON	&	2456677.76	&	2014.0520	&	12.90278	&	2.83	&	1.73	&	2.53	&	1.64	\\
204	&	CTIO/CHIRON	&	2456687.71	&	2014.0793	&	12.90770	&	2.85	&	1.79	&	2.55	&	1.64	\\
205	&	CTIO/CHIRON	&	2456690.69	&	2014.0875	&	12.90918	&	2.85	&	1.79	&	2.62	&	1.64	\\
206	&	CTIO/CHIRON	&	2456697.73	&	2014.1067	&	12.91265	&	2.83	&	1.74	&	2.53	&	1.64	\\
207	&	CTIO/CHIRON	&	2456710.67	&	2014.1421	&	12.91905	&	2.80	&	1.69	&	2.37	&	1.64	\\
208	&	CTIO/CHIRON	&	2456712.69	&	2014.1477	&	12.92005	&	2.81	&	1.77	&	2.50	&	1.64	\\
209	&	CTIO/CHIRON	&	2456718.71	&	2014.1642	&	12.92303	&	2.80	&	1.73	&	2.40	&	1.64	\\
210	&	CTIO/CHIRON	&	2456725.57	&	2014.1829	&	12.92642	&	2.90	&	1.82	&	2.51	&	1.64	\\
211	&	CTIO/CHIRON	&	2456729.56	&	2014.1939	&	12.92839	&	2.81	&	1.77	&	2.44	&	1.64	\\
212	&	CTIO/CHIRON	&	2456732.61	&	2014.2022	&	12.92990	&	2.95	&	1.76	&	2.47	&	1.64	\\
213	&	CTIO/CHIRON	&	2456739.55	&	2014.2212	&	12.93333	&	2.92	&	1.74	&	2.29	&	1.64	\\
214	&	CTIO/CHIRON	&	2456746.52	&	2014.2403	&	12.93678	&	2.90	&	1.82	&	2.40	&	1.64	\\
215	&	CTIO/CHIRON	&	2456750.54	&	2014.2513	&	12.93877	&	2.90	&	1.80	&	2.30	&	1.64	\\
216	&	CTIO/CHIRON	&	2456754.55	&	2014.2623	&	12.94075	&	2.88	&	1.80	&	2.23	&	1.64	\\
217	&	CTIO/CHIRON	&	2456765.59	&	2014.2925	&	12.94621	&	2.94	&	1.83	&	2.20	&	1.64	\\
218	&	CTIO/CHIRON	&	2456766.55	&	2014.2951	&	12.94668	&	2.82	&	1.78	&	2.13	&	1.64	\\
219	&	CTIO/CHIRON	&	2456767.51	&	2014.2978	&	12.94716	&	2.76	&	1.78	&	2.02	&	1.64	\\
220	&	CTIO/CHIRON	&	2456774.55	&	2014.3170	&	12.95063	&	2.73	&	1.77	&	1.98	&	1.64	\\
221	&	CTIO/CHIRON	&	2456781.47	&	2014.3360	&	12.95406	&	2.69	&	1.74	&	1.95	&	1.64	\\
222	&	CTIO/CHIRON	&	2456791.55	&	2014.3636	&	12.95904	&	2.57	&	1.71	&	1.55	&	1.76	\\
223	&	CTIO/CHIRON	&	2456795.51	&	2014.3744	&	12.96100	&	2.49	&	1.70	&	1.39	&	1.76	\\
224	&	CTIO/CHIRON	&	2456800.46	&	2014.3880	&	12.96345	&	2.34	&	1.69	&	1.21	&	1.76	\\
225	&	CTIO/CHIRON	&	2456801.50	&	2014.3908	&	12.96396	&	2.23	&	1.65	&	1.15	&	1.76	\\
226	&	CTIO/CHIRON	&	2456803.53	&	2014.3964	&	12.96496	&	1.99	&	1.63	&	1.09	&	1.76	\\
227	&	CTIO/CHIRON	&	2456810.48	&	2014.4154	&	12.96840	&	1.98	&	1.61	&	1.09	&	1.76	\\
228	&	CTIO/CHIRON	&	2456818.51	&	2014.4374	&	12.97237	&	1.93	&	1.58	&	1.08	&	1.76	\\
229	&	CTIO/CHIRON	&	2456819.53	&	2014.4402	&	12.97287	&	1.89	&	1.59	&	1.08	&	1.76	\\
230	&	CTIO/CHIRON	&	2456823.54	&	2014.4512	&	12.97486	&	1.90	&	1.60	&	1.08	&	1.95	\\
231	&	CTIO/CHIRON	&	2456824.52	&	2014.4538	&	12.97534	&	1.84	&	1.58	&	1.09	&	1.95	\\
232	&	CTIO/CHIRON	&	2456829.50	&	2014.4675	&	12.97780	&	1.83	&	1.55	&	1.08	&	1.95	\\
233	&	CTIO/CHIRON	&	2456832.50	&	2014.4757	&	12.97929	&	1.73	&	1.57	&	1.10	&	1.95	\\
234	&	CTIO/CHIRON	&	2456835.53	&	2014.4840	&	12.98079	&	1.75	&	1.56	&	1.10	&	1.95	\\
235	&	CTIO/CHIRON	&	2456836.50	&	2014.4866	&	12.98126	&	1.75	&	1.54	&	1.08	&	1.95	\\
236	&	CTIO/CHIRON	&	2456845.46	&	2014.5112	&	12.98569	&	1.74	&	1.57	&	1.09	&	1.95	\\
237	&	CTIO/CHIRON	&	2456850.51	&	2014.5250	&	12.98819	&	2.20	&	1.57	&	1.07	&	1.95	\\
238	&	CTIO/CHIRON	&	2456855.46	&	2014.5386	&	12.99064	&	2.15	&	1.54	&	1.20	&	1.95	\\
239	&	CTIO/CHIRON	&	2456857.45	&	2014.5440	&	12.99162	&	2.29	&	1.57	&	1.23	&	1.95	\\
240	&	CTIO/CHIRON	&	2456858.45	&	2014.5468	&	12.99212	&	2.30	&	1.62	&	1.24	&	1.95	\\
241	&	CTIO/CHIRON	&	2456859.47	&	2014.5495	&	12.99262	&	2.30	&	1.56	&	1.22	&	1.95	\\
242	&	CTIO/CHIRON	&	2456863.47	&	2014.5605	&	12.99459	&	2.32	&	1.57	&	1.25	&	1.91	\\
243	&	CTIO/CHIRON	&	2456864.47	&	2014.5632	&	12.99509	&	2.36	&	1.59	&	1.23	&	1.91	\\
244	&	CTIO/CHIRON	&	2456866.51	&	2014.5688	&	12.99610	&	2.30	&	1.58	&	1.27	&	1.91	\\
245	&	CTIO/CHIRON	&	2456867.46	&	2014.5714	&	12.99657	&	2.35	&	1.59	&	1.29	&	1.91	\\
246	&	CTIO/CHIRON	&	2456870.46	&	2014.5796	&	12.99805	&	2.27	&	1.58	&	1.26	&	1.91	\\
247	&	CTIO/CHIRON	&	2456871.46	&	2014.5824	&	12.99855	&	2.31	&	1.59	&	1.28	&	1.91	\\
248	&	CTIO/CHIRON	&	2456872.46	&	2014.5851	&	12.99904	&	2.37	&	1.61	&	1.32	&	1.91	\\
249	&	CTIO/CHIRON	&	2456873.46	&	2014.5879	&	12.99954	&	2.39	&	1.62	&	1.34	&	1.91	\\
250	&	CTIO/CHIRON	&	2456874.45	&	2014.5906	&	13.00003	&	2.34	&	1.63	&	1.37	&	1.91	\\
251	&	CTIO/CHIRON	&	2456878.48	&	2014.6016	&	13.00202	&	2.37	&	1.58	&	1.35	&	1.91	\\
252	&	CTIO/CHIRON	&	2456879.47	&	2014.6043	&	13.00250	&	2.35	&	1.60	&	1.38	&	1.91	\\
253	&	CTIO/CHIRON	&	2456882.46	&	2014.6125	&	13.00399	&	2.33	&	1.55	&	1.36	&	1.91	\\
254	&	CTIO/CHIRON	&	2456883.46	&	2014.6152	&	13.00448	&	2.23	&	1.52	&	1.37	&	1.91	\\
255	&	CTIO/CHIRON	&	2456885.46	&	2014.6207	&	13.00547	&	2.31	&	1.56	&	1.41	&	1.91	\\
256	&	CTIO/CHIRON	&	2456886.47	&	2014.6235	&	13.00597	&	2.31	&	1.55	&	1.43	&	1.91	\\
257	&	CTIO/CHIRON	&	2456943.90	&	2014.7807	&	13.03436	&	2.36	&	1.59	&	1.48	&	1.78	\\
258	&	CTIO/CHIRON	&	2456944.90	&	2014.7834	&	13.03485	&	2.34	&	1.59	&	1.46	&	1.78	\\
259	&	CTIO/CHIRON	&	2456950.88	&	2014.7998	&	13.03781	&	2.27	&	1.54	&	1.45	&	1.78	\\
260	&	CTIO/CHIRON	&	2456951.87	&	2014.8025	&	13.03830	&	2.33	&	1.57	&	1.48	&	1.78	\\
261	&	CTIO/CHIRON	&	2456952.86	&	2014.8052	&	13.03879	&	2.26	&	1.53	&	1.47	&	1.78	\\
262	&	CTIO/CHIRON	&	2456953.88	&	2014.8080	&	13.03929	&	2.26	&	1.52	&	1.49	&	1.78	\\
263	&	CTIO/CHIRON	&	2456954.83	&	2014.8106	&	13.03976	&	2.31	&	1.56	&	1.49	&	1.78	\\
264	&	CTIO/CHIRON	&	2456955.86	&	2014.8134	&	13.04027	&	2.36	&	1.59	&	1.54	&	1.78	\\
265	&	CTIO/CHIRON	&	2456956.87	&	2014.8162	&	13.04077	&	2.34	&	1.59	&	1.52	&	1.78	\\
266	&	CTIO/CHIRON	&	2456957.89	&	2014.8190	&	13.04128	&	2.33	&	1.55	&	1.53	&	1.78	\\
267	&	CTIO/CHIRON	&	2456958.86	&	2014.8217	&	13.04176	&	2.27	&	1.53	&	1.56	&	1.78	\\
268	&	CTIO/CHIRON	&	2456959.85	&	2014.8244	&	13.04224	&	2.26	&	1.53	&	1.56	&	1.78	\\
269	&	CTIO/CHIRON	&	2456961.85	&	2014.8298	&	13.04323	&	2.29	&	1.54	&	1.54	&	1.78	\\
270	&	CTIO/CHIRON	&	2456964.87	&	2014.8381	&	13.04472	&	2.34	&	1.54	&	1.62	&	1.78	\\
271	&	CTIO/CHIRON	&	2456965.86	&	2014.8408	&	13.04522	&	2.30	&	1.54	&	1.60	&	1.78	\\
272	&	CTIO/CHIRON	&	2456968.87	&	2014.8491	&	13.04670	&	2.30	&	1.53	&	1.61	&	1.78	\\
273	&	CTIO/CHIRON	&	2456969.86	&	2014.8518	&	13.04720	&	2.27	&	1.53	&	1.59	&	1.78	\\
274	&	CTIO/CHIRON	&	2456972.88	&	2014.8600	&	13.04869	&	2.31	&	1.55	&	1.65	&	1.78	\\
275	&	CTIO/CHIRON	&	2456974.85	&	2014.8654	&	13.04966	&	2.29	&	1.55	&	1.62	&	1.78	\\
276	&	CTIO/CHIRON	&	2456977.85	&	2014.8736	&	13.05114	&	2.23	&	1.53	&	1.61	&	1.78	\\
277	&	CTIO/CHIRON	&	2456979.84	&	2014.8791	&	13.05213	&	2.25	&	1.52	&	1.63	&	1.78	\\
278	&	CTIO/CHIRON	&	2456980.81	&	2014.8817	&	13.05261	&	2.30	&	1.54	&	1.66	&	1.78	\\
279	&	CTIO/CHIRON	&	2456982.81	&	2014.8872	&	13.05360	&	2.23	&	1.54	&	1.65	&	1.78	\\
280	&	CTIO/CHIRON	&	2456984.82	&	2014.8927	&	13.05459	&	2.29	&	1.58	&	1.69	&	1.78	\\
281	&	CTIO/CHIRON	&	2456985.83	&	2014.8955	&	13.05509	&	2.18	&	1.54	&	1.64	&	1.78	\\
282	&	CTIO/CHIRON	&	2456986.87	&	2014.8983	&	13.05560	&	2.25	&	1.51	&	1.67	&	1.78	\\
283	&	CTIO/CHIRON	&	2456988.79	&	2014.9036	&	13.05656	&	2.16	&	1.55	&	1.70	&	1.78	\\
284	&	CTIO/CHIRON	&	2456990.76	&	2014.9090	&	13.05753	&	2.23	&	1.54	&	1.73	&	1.78	\\
285	&	CTIO/CHIRON	&	2456992.78	&	2014.9145	&	13.05853	&	2.22	&	1.51	&	1.70	&	1.78	\\
286	&	CTIO/CHIRON	&	2456994.82	&	2014.9201	&	13.05953	&	2.26	&	1.49	&	1.70	&	1.78	\\
287	&	CTIO/CHIRON	&	2456996.85	&	2014.9257	&	13.06054	&	2.25	&	1.56	&	1.74	&	1.78	\\
288	&	CTIO/CHIRON	&	2456998.81	&	2014.9310	&	13.06150	&	2.21	&	1.52	&	1.70	&	1.78	\\
289	&	CTIO/CHIRON	&	2457000.84	&	2014.9366	&	13.06251	&	2.23	&	1.52	&	1.73	&	1.78	\\
290	&	CTIO/CHIRON	&	2457007.85	&	2014.9558	&	13.06598	&	2.23	&	1.52	&	1.71	&	1.83	\\
291	&	CTIO/CHIRON	&	2457008.74	&	2014.9582	&	13.06642	&	2.25	&	1.54	&	1.75	&	1.83	\\
292	&	CTIO/CHIRON	&	2457009.75	&	2014.9610	&	13.06691	&	2.16	&	1.47	&	1.68	&	1.83	\\
293	&	CTIO/CHIRON	&	2457012.82	&	2014.9694	&	13.06843	&	2.25	&	1.54	&	1.74	&	1.83	\\
294	&	CTIO/CHIRON	&	2457013.81	&	2014.9721	&	13.06892	&	2.22	&	1.54	&	1.78	&	1.83	\\
295	&	CTIO/CHIRON	&	2457014.74	&	2014.9746	&	13.06938	&	2.28	&	1.53	&	1.80	&	1.83	\\
296	&	CTIO/CHIRON	&	2457015.77	&	2014.9775	&	13.06989	&	2.19	&	1.51	&	1.77	&	1.83	\\
297	&	CTIO/CHIRON	&	2457016.80	&	2014.9803	&	13.07040	&	2.09	&	1.46	&	1.73	&	1.83	\\
298	&	CTIO/CHIRON	&	2457018.78	&	2014.9857	&	13.07138	&	2.23	&	1.52	&	1.84	&	1.83	\\
299	&	CTIO/CHIRON	&	2457021.84	&	2014.9941	&	13.07289	&	2.14	&	1.45	&	1.77	&	1.83	\\
300	&	CTIO/CHIRON	&	2457022.75	&	2014.9966	&	13.07334	&	2.15	&	1.51	&	1.77	&	1.83	\\
301	&	CTIO/CHIRON	&	2457025.83	&	2015.0050	&	13.07486	&	2.12	&	1.48	&	1.77	&	1.83	\\
302	&	CTIO/CHIRON	&	2457026.70	&	2015.0074	&	13.07530	&	2.11	&	1.49	&	1.72	&	1.83	\\
303	&	CTIO/CHIRON	&	2457030.69	&	2015.0183	&	13.07727	&	2.06	&	1.45	&	1.71	&	1.83	\\
304	&	CTIO/CHIRON	&	2457032.82	&	2015.0241	&	13.07832	&	2.27	&	1.56	&	1.91	&	1.83	\\
305	&	CTIO/CHIRON	&	2457035.72	&	2015.0321	&	13.07976	&	2.15	&	1.51	&	1.80	&	1.83	\\
306	&	CTIO/CHIRON	&	2457037.82	&	2015.0378	&	13.08079	&	2.25	&	1.54	&	1.95	&	1.83	\\
307	&	CTIO/CHIRON	&	2457038.85	&	2015.0407	&	13.08130	&	2.04	&	1.42	&	1.74	&	1.83	\\
308	&	CTIO/CHIRON	&	2457039.83	&	2015.0434	&	13.08179	&	2.06	&	1.45	&	1.81	&	1.83	\\
309	&	CTIO/CHIRON	&	2457040.71	&	2015.0458	&	13.08222	&	2.07	&	1.45	&	1.83	&	1.83	\\
310	&	CTIO/CHIRON	&	2457043.85	&	2015.0543	&	13.08377	&	2.15	&	1.50	&	1.87	&	1.83	\\
311	&	CTIO/CHIRON	&	2457046.68	&	2015.0621	&	13.08518	&	2.14	&	1.51	&	1.90	&	1.83	\\
312	&	CTIO/CHIRON	&	2457047.69	&	2015.0648	&	13.08567	&	2.18	&	1.49	&	1.92	&	1.83	\\
313	&	CTIO/CHIRON	&	2457048.81	&	2015.0679	&	13.08623	&	2.20	&	1.49	&	1.92	&	1.83	\\
314	&	CTIO/CHIRON	&	2457050.73	&	2015.0732	&	13.08718	&	2.08	&	1.45	&	1.84	&	1.83	\\
315	&	CTIO/CHIRON	&	2457051.85	&	2015.0762	&	13.08773	&	2.20	&	1.50	&	1.94	&	1.83	\\
316	&	CTIO/CHIRON	&	2457059.79	&	2015.0980	&	13.09165	&	2.16	&	1.51	&	1.94	&	1.83	\\
317	&	CTIO/CHIRON	&	2457060.66	&	2015.1004	&	13.09209	&	2.14	&	1.49	&	1.91	&	1.83	\\
318	&	CTIO/CHIRON	&	2457061.85	&	2015.1036	&	13.09267	&	2.09	&	1.45	&	1.85	&	1.83	\\
319	&	CTIO/CHIRON	&	2457062.70	&	2015.1059	&	13.09309	&	2.18	&	1.51	&	1.95	&	1.83	\\
320	&	CTIO/CHIRON	&	2457063.74	&	2015.1088	&	13.09361	&	2.11	&	1.46	&	1.88	&	1.83	\\
321	&	CTIO/CHIRON	&	2457064.78	&	2015.1116	&	13.09412	&	2.07	&	1.42	&	1.90	&	1.83	\\
322	&	CTIO/CHIRON	&	2457068.77	&	2015.1226	&	13.09609	&	2.15	&	1.49	&	1.92	&	1.83	\\
323	&	CTIO/CHIRON	&	2457070.65	&	2015.1277	&	13.09702	&	2.03	&	1.43	&	1.89	&	1.83	\\
324	&	CTIO/CHIRON	&	2457074.80	&	2015.1391	&	13.09908	&	2.10	&	1.48	&	1.94	&	1.83	\\
325	&	CTIO/CHIRON	&	2457075.70	&	2015.1415	&	13.09952	&	2.06	&	1.43	&	1.88	&	1.83	\\
326	&	CTIO/CHIRON	&	2457078.90	&	2015.1503	&	13.10110	&	2.04	&	1.42	&	1.91	&	1.83	\\
327	&	CTIO/CHIRON	&	2457083.66	&	2015.1633	&	13.10346	&	2.05	&	1.45	&	1.94	&	1.83	\\
328	&	CTIO/CHIRON	&	2457088.64	&	2015.1770	&	13.10592	&	2.07	&	1.44	&	1.97	&	1.83	\\
329	&	CTIO/CHIRON	&	2457091.59	&	2015.1851	&	13.10738	&	2.06	&	1.45	&	1.97	&	1.83	\\
330	&	CTIO/CHIRON	&	2457092.72	&	2015.1882	&	13.10794	&	2.06	&	1.42	&	2.00	&	1.83	\\
331	&	CTIO/CHIRON	&	2457093.71	&	2015.1909	&	13.10842	&	2.08	&	1.46	&	1.99	&	1.83	\\
332	&	CTIO/CHIRON	&	2457095.80	&	2015.1966	&	13.10946	&	2.06	&	1.44	&	1.95	&	1.83	\\
333	&	CTIO/CHIRON	&	2457097.70	&	2015.2018	&	13.11040	&	2.05	&	1.42	&	1.95	&	1.83	\\
334	&	CTIO/CHIRON	&	2457104.56	&	2015.2206	&	13.11379	&	2.14	&	1.43	&	2.03	&	1.83	\\
335	&	CTIO/CHIRON	&	2457115.75	&	2015.2512	&	13.11932	&	2.16	&	1.49	&	1.99	&	1.83	\\
336	&	CTIO/CHIRON	&	2457118.56	&	2015.2589	&	13.12071	&	2.11	&	1.47	&	1.98	&	1.83	\\
337	&	CTIO/CHIRON	&	2457120.54	&	2015.2643	&	13.12169	&	2.12	&	1.46	&	2.03	&	1.83	\\
338	&	CTIO/CHIRON	&	2457124.50	&	2015.2752	&	13.12365	&	2.04	&	1.46	&	2.31	&	1.83	\\
339	&	CTIO/CHIRON	&	2457125.52	&	2015.2780	&	13.12415	&	2.16	&	1.43	&	2.36	&	1.83	\\
340	&	CTIO/CHIRON	&	2457127.62	&	2015.2837	&	13.12519	&	2.41	&	1.58	&	2.30	&	1.83	\\
341	&	CTIO/CHIRON	&	2457132.52	&	2015.2971	&	13.12761	&	2.39	&	1.55	&	2.25	&	1.83	\\
342	&	CTIO/CHIRON	&	2457134.63	&	2015.3029	&	13.12865	&	2.37	&	1.54	&	2.21	&	1.83	\\
343	&	CTIO/CHIRON	&	2457145.63	&	2015.3330	&	13.13409	&	2.38	&	1.55	&	2.15	&	1.83	\\
344	&	CTIO/CHIRON	&	2457150.62	&	2015.3467	&	13.13656	&	2.41	&	1.56	&	2.25	&	1.83	\\
345	&	CTIO/CHIRON	&	2457155.57	&	2015.3602	&	13.13901	&	2.34	&	1.53	&	2.13	&	1.83	\\
346	&	CTIO/CHIRON	&	2457161.55	&	2015.3766	&	13.14197	&	2.50	&	1.62	&	2.31	&	1.83	\\
347	&	CTIO/CHIRON	&	2457164.55	&	2015.3848	&	13.14345	&	2.49	&	1.62	&	2.27	&	1.83	\\
348	&	CTIO/CHIRON	&	2457166.53	&	2015.3902	&	13.14443	&	2.44	&	1.60	&	2.22	&	1.83	\\
349	&	CTIO/CHIRON	&	2457172.52	&	2015.4066	&	13.14739	&	2.50	&	1.62	&	2.28	&	1.83	\\
350	&	CTIO/CHIRON	&	2457174.59	&	2015.4123	&	13.14841	&	2.36	&	1.55	&	2.33	&	1.83	\\
351	&	CTIO/CHIRON	&	2457177.49	&	2015.4202	&	13.14984	&	2.46	&	1.58	&	2.22	&	1.83	\\
352	&	CTIO/CHIRON	&	2457181.45	&	2015.4311	&	13.15180	&	2.50	&	1.61	&	2.25	&	1.83	\\
353	&	CTIO/CHIRON	&	2457185.50	&	2015.4422	&	13.15381	&	2.65	&	1.67	&	2.28	&	1.83	\\
354	&	CTIO/CHIRON	&	2457189.50	&	2015.4531	&	13.15578	&	2.55	&	1.67	&	2.46	&	1.83	\\
355	&	CTIO/CHIRON	&	2457191.46	&	2015.4585	&	13.15675	&	2.56	&	1.63	&	2.17	&	2.06	\\
356	&	CTIO/CHIRON	&	2457194.50	&	2015.4668	&	13.15825	&	2.47	&	1.61	&	2.42	&	2.06	\\
357	&	CTIO/CHIRON	&	2457212.44	&	2015.5159	&	13.16713	&	2.61	&	1.65	&	2.23	&	2.06	\\
358	&	ESO/FEROS	&	2457732.84	&	2016.9407	&	13.42440	&	1.40	&	1.30	&	1.02	&	2.06	\\
359	&	LCO/NRES	&	2458502.44	&	2019.0478	&	13.80489	&	4.61	&	2.59	&	3.16	&	1.99	\\
360	&	CTIO/CHIRON	&	2458777.87	&	2019.8018	&	13.94105	&	2.63	&	1.66	&	2.34	&	2.24	\\
361	&	CTIO/CHIRON	&	2458782.88	&	2019.8156	&	13.94353	&	2.49	&	1.66	&	2.41	&	2.24	\\
362	&	CTIO/CHIRON	&	2458785.87	&	2019.8237	&	13.94501	&	2.54	&	1.60	&	2.40	&	2.24	\\
363	&	CTIO/CHIRON	&	2458800.86	&	2019.8648	&	13.95242	&	2.49	&	1.71	&	2.50	&	2.24	\\
364	&	CTIO/CHIRON	&	2458803.85	&	2019.8730	&	13.95390	&	2.50	&	1.69	&	2.34	&	2.24	\\
365	&	CTIO/CHIRON	&	2458805.83	&	2019.8784	&	13.95488	&	2.37	&	1.60	&	2.16	&	2.24	\\
366	&	CTIO/CHIRON	&	2458809.86	&	2019.8894	&	13.95687	&	2.54	&	1.63	&	2.18	&	2.24	\\
367	&	CTIO/CHIRON	&	2458812.86	&	2019.8976	&	13.95835	&	2.45	&	1.60	&	2.20	&	2.24	\\
368	&	CTIO/CHIRON	&	2458814.86	&	2019.9031	&	13.95934	&	2.44	&	1.60	&	2.21	&	2.24	\\
369	&	CTIO/CHIRON	&	2458817.86	&	2019.9113	&	13.96083	&	2.46	&	1.70	&	2.44	&	2.24	\\
370	&	LCO/NRES	&	2458820.80	&	2019.9194	&	13.96228	&	4.02	&	2.45	&	2.19	&	2.24	\\
371	&	CTIO/CHIRON	&	2458820.81	&	2019.9194	&	13.96228	&	2.46	&	1.60	&	2.33	&	2.24	\\
372	&	CTIO/CHIRON	&	2458822.81	&	2019.9249	&	13.96327	&	2.49	&	1.74	&	2.46	&	2.24	\\
373	&	CTIO/CHIRON	&	2458824.79	&	2019.9303	&	13.96425	&	2.49	&	1.69	&	2.34	&	2.24	\\
374	&	LCO/NRES	&	2458827.79	&	2019.9385	&	13.96573	&	3.85	&	2.41	&	2.25	&	2.24	\\
375	&	CTIO/CHIRON	&	2458827.86	&	2019.9387	&	13.96577	&	2.28	&	1.53	&	2.18	&	2.24	\\
376	&	LCO/NRES	&	2458829.75	&	2019.9439	&	13.96670	&	3.84	&	2.41	&	2.21	&	2.24	\\
377	&	CTIO/CHIRON	&	2458830.82	&	2019.9468	&	13.96723	&	2.28	&	1.53	&	2.40	&	2.24	\\
378	&	CTIO/CHIRON	&	2458832.86	&	2019.9524	&	13.96824	&	2.07	&	1.45	&	2.02	&	2.20	\\
379	&	LCO/NRES	&	2458833.79	&	2019.9549	&	13.96870	&	3.77	&	2.40	&	2.19	&	2.20	\\
380	&	CTIO/CHIRON	&	2458834.85	&	2019.9578	&	13.96922	&	2.09	&	1.45	&	2.01	&	2.20	\\
381	&	LCO/NRES	&	2458836.81	&	2019.9632	&	13.97019	&	4.50	&	2.44	&	2.20	&	2.20	\\
382	&	CTIO/CHIRON	&	2458836.82	&	2019.9632	&	13.97020	&	2.16	&	1.49	&	2.08	&	2.20	\\
383	&	LCO/NRES	&	2458839.52	&	2019.9706	&	13.97153	&	4.72	&	2.88	&	1.36	&	2.20	\\
384	&	LCO/NRES	&	2458841.49	&	2019.9760	&	13.97251	&	3.30	&	2.45	&	1.22	&	2.20	\\
385	&	LCO/NRES	&	2458843.50	&	2019.9815	&	13.97350	&	3.29	&	2.45	&	1.18	&	2.20	\\
386	&	LCO/NRES	&	2458845.52	&	2019.9870	&	13.97450	&	4.15	&	2.77	&	1.36	&	2.20	\\
387	&	LCO/NRES	&	2458847.49	&	2019.9925	&	13.97547	&	4.43	&	2.77	&	1.33	&	2.20	\\
388	&	LCO/NRES	&	2458848.72	&	2019.9958	&	13.97608	&	3.68	&	2.26	&	1.69	&	2.20	\\
389	&	LCO/NRES	&	2458851.75	&	2020.0041	&	13.97758	&	3.68	&	2.26	&	1.69	&	2.20	\\
390	&	LCO/NRES	&	2458854.79	&	2020.0124	&	13.97908	&	4.85	&	2.72	&	3.26	&	2.20	\\
391	&	LCO/NRES	&	2458857.78	&	2020.0206	&	13.98056	&	4.90	&	2.56	&	3.20	&	2.20	\\
392	&	LCO/NRES	&	2458858.71	&	2020.0232	&	13.98102	&	5.28	&	2.63	&	3.40	&	2.20	\\
393	&	LCO/NRES	&	2458859.71	&	2020.0259	&	13.98152	&	5.01	&	2.54	&	3.29	&	2.20	\\
394	&	LCO/NRES	&	2458860.85	&	2020.0290	&	13.98208	&	3.65	&	2.29	&	3.04	&	2.20	\\
395	&	LCO/NRES	&	2458861.81	&	2020.0316	&	13.98255	&	3.61	&	2.26	&	3.00	&	2.20	\\
396	&	LCO/NRES	&	2458862.75	&	2020.0342	&	13.98302	&	4.54	&	2.31	&	3.49	&	2.20	\\
397	&	LCO/NRES	&	2458863.70	&	2020.0368	&	13.98349	&	6.29	&	3.03	&	4.76	&	2.20	\\
398	&	LCO/NRES	&	2458864.83	&	2020.0399	&	13.98405	&	7.00	&	3.17	&	4.70	&	2.20	\\
399	&	LCO/NRES	&	2458866.81	&	2020.0453	&	13.98503	&	7.15	&	3.22	&	4.75	&	2.20	\\
400	&	LCO/NRES	&	2458869.84	&	2020.0536	&	13.98652	&	7.72	&	3.45	&	5.19	&	2.62	\\
401	&	LCO/NRES	&	2458871.78	&	2020.0590	&	13.98748	&	8.29	&	3.49	&	5.29	&	2.62	\\
402	&	LCO/NRES	&	2458872.86	&	2020.0619	&	13.98802	&	3.50	&	3.70	&	3.30	&	2.62	\\
403	&	LCO/NRES	&	2458874.38	&	2020.0661	&	13.98877	&	3.00	&	3.11	&	3.20	&	2.62	\\
404	&	LCO/NRES	&	2458875.80	&	2020.0700	&	13.98947	&	2.72	&	1.91	&	1.73	&	2.62	\\
405	&	LCO/NRES	&	2458876.85	&	2020.0728	&	13.98999	&	2.65	&	1.81	&	1.58	&	2.62	\\
406	&	LCO/NRES	&	2458877.50	&	2020.0746	&	13.99031	&	2.67	&	1.85	&	1.49	&	2.62	\\
407	&	LCO/NRES	&	2458878.54	&	2020.0774	&	13.99082	&	2.76	&	1.93	&	1.46	&	2.62	\\
408	&	LCO/NRES	&	2458879.75	&	2020.0808	&	13.99142	&	2.84	&	2.03	&	1.44	&	2.62	\\
409	&	LCO/NRES	&	2458880.84	&	2020.0838	&	13.99196	&	2.72	&	1.98	&	1.43	&	2.62	\\
410	&	LCO/NRES	&	2458882.83	&	2020.0892	&	13.99295	&	2.83	&	2.09	&	1.38	&	2.62	\\
411	&	LCO/NRES	&	2458882.83	&	2020.0892	&	13.99295	&	2.81	&	2.06	&	1.36	&	2.62	\\
412	&	LCO/NRES	&	2458884.58	&	2020.0940	&	13.99381	&	2.28	&	1.76	&	1.09	&	2.62	\\
413	&	LCO/NRES	&	2458884.60	&	2020.0940	&	13.99382	&	2.57	&	1.93	&	1.28	&	2.62	\\
414	&	LCO/NRES	&	2458885.68	&	2020.0970	&	13.99436	&	2.09	&	1.70	&	1.11	&	2.62	\\
415	&	LCO/NRES	&	2458886.82	&	2020.1001	&	13.99492	&	1.90	&	1.60	&	1.08	&	2.62	\\
416	&	LCO/NRES	&	2458887.61	&	2020.1023	&	13.99531	&	1.93	&	1.64	&	1.08	&	2.62	\\
417	&	LCO/NRES	&	2458888.53	&	2020.1048	&	13.99576	&	1.88	&	1.59	&	1.05	&	2.62	\\
418	&	LCO/NRES	&	2458888.67	&	2020.1052	&	13.99583	&	1.91	&	1.65	&	1.09	&	2.62	\\
419	&	LCO/NRES	&	2458889.75	&	2020.1081	&	13.99636	&	1.94	&	1.66	&	1.09	&	2.62	\\
420	&	LCO/NRES	&	2458890.78	&	2020.1110	&	13.99688	&	2.12	&	1.76	&	1.08	&	2.62	\\
421	&	LCO/NRES	&	2458891.64	&	2020.1133	&	13.99730	&	2.03	&	1.76	&	1.09	&	2.62	\\
422	&	LCO/NRES	&	2458892.80	&	2020.1165	&	13.99788	&	2.05	&	1.77	&	1.10	&	2.62	\\
423	&	LCO/NRES	&	2458893.50	&	2020.1184	&	13.99822	&	2.00	&	1.75	&	1.17	&	2.62	\\
424	&	LCO/NRES	&	2458893.79	&	2020.1192	&	13.99836	&	1.97	&	1.76	&	1.12	&	2.62	\\
425	&	LCO/NRES	&	2458894.80	&	2020.1220	&	13.99886	&	1.94	&	1.75	&	1.10	&	2.62	\\
426	&	LCO/NRES	&	2458896.54	&	2020.1267	&	13.99972	&	1.94	&	1.73	&	1.09	&	2.62	\\
427	&	LCO/NRES	&	2458896.85	&	2020.1276	&	13.99988	&	1.88	&	1.65	&	1.09	&	2.62	\\
428	&	LCO/NRES	&	2458898.39	&	2020.1318	&	14.00064	&	1.84	&	1.68	&	1.10	&	2.62	\\
429	&	LCO/NRES	&	2458899.79	&	2020.1356	&	14.00133	&	1.79	&	1.64	&	1.07	&	2.62	\\
430	&	LCO/NRES	&	2458900.78	&	2020.1383	&	14.00182	&	1.84	&	1.69	&	1.09	&	2.62	\\
431	&	LCO/NRES	&	2458903.71	&	2020.1464	&	14.00327	&	1.77	&	1.63	&	1.07	&	2.62	\\
432	&	LCO/NRES	&	2458904.70	&	2020.1491	&	14.00376	&	1.87	&	1.65	&	1.09	&	2.62	\\
433	&	LCO/NRES	&	2458906.79	&	2020.1548	&	14.00479	&	2.02	&	1.59	&	1.34	&	2.24	\\
434	&	LCO/NRES	&	2458908.72	&	2020.1601	&	14.00574	&	2.42	&	1.67	&	1.30	&	2.24	\\
435	&	LCO/NRES	&	2458910.61	&	2020.1653	&	14.00668	&	3.72	&	2.18	&	2.89	&	2.24	\\
436	&	LCO/NRES	&	2458912.81	&	2020.1713	&	14.00777	&	3.12	&	1.96	&	2.82	&	2.24	\\
437	&	LCO/NRES	&	2458914.63	&	2020.1763	&	14.00867	&	5.77	&	2.75	&	4.34	&	2.24	\\
438	&	LCO/NRES	&	2458916.66	&	2020.1818	&	14.00967	&	5.60	&	2.70	&	4.14	&	2.24	\\
439	&	LCO/NRES	&	2458918.68	&	2020.1873	&	14.01067	&	5.50	&	2.76	&	4.10	&	2.24	\\
440	&	LCO/NRES	&	2458920.73	&	2020.1930	&	14.01168	&	5.86	&	2.76	&	4.17	&	2.24	\\
441	&	LCO/NRES	&	2458922.73	&	2020.1984	&	14.01267	&	5.59	&	2.74	&	3.96	&	2.24	\\
442	&	LCO/NRES	&	2458924.65	&	2020.2037	&	14.01362	&	5.80	&	2.67	&	4.03	&	2.24	\\
443	&	LCO/NRES	&	2458926.52	&	2020.2088	&	14.01454	&	5.40	&	2.67	&	4.03	&	2.24	\\
444	&	LCO/NRES	&	2458927.50	&	2020.2115	&	14.01503	&	5.50	&	2.86	&	4.00	&	2.24	\\
445	&	LCO/NRES	&	2458928.48	&	2020.2142	&	14.01551	&	5.60	&	2.76	&	4.00	&	2.24	\\
446	&	LCO/NRES	&	2458930.46	&	2020.2196	&	14.01649	&	5.46	&	2.75	&	3.91	&	2.24	\\
447	&	LCO/NRES	&	2458931.26	&	2020.2218	&	14.01689	&	5.39	&	2.48	&	3.77	&	2.24	\\
448	&	LCO/NRES	&	2458991.20	&	2020.3859	&	14.04652	&	5.15	&	2.61	&	3.48	&	2.24	\\
449	&	LCO/NRES	&	2459000.20	&	2020.4105	&	14.05097	&	4.25	&	2.33	&	2.32	&	2.48	\\
450	&	LCO/NRES	&	2459035.24	&	2020.5065	&	14.06829	&	4.13	&	2.27	&	2.17	&	2.59	\\
451	&	LCO/NRES	&	2459039.23	&	2020.5174	&	14.07027	&	4.12	&	2.26	&	2.11	&	2.59	\\
452	&	LCO/NRES	&	2459051.20	&	2020.5502	&	14.07619	&	4.06	&	2.26	&	2.07	&	2.57	\\
453	&	LCO/NRES	&	2459152.59	&	2020.8278	&	14.12631	&	4.10	&	2.18	&	1.96	&	2.57	\\
454	&	LCO/NRES	&	2459155.59	&	2020.8360	&	14.12780	&	4.08	&	2.24	&	1.92	&	2.57	\\
455	&	LCO/NRES	&	2459252.61	&	2021.1016	&	14.17576	&	4.12	&	2.33	&	2.00	&	2.64	\\
456	&	LCO/NRES	&	2459265.54	&	2021.1370	&	14.18215	&	3.80	&	2.16	&	1.79	&	2.64	\\
457	&	LCO/NRES	&	2459277.44	&	2021.1696	&	14.18804	&	3.88	&	2.27	&	1.84	&	2.64	\\
458	&	LCO/NRES	&	2459290.38	&	2021.2050	&	14.19444	&	3.47	&	2.04	&	1.74	&	2.64	\\
459	&	LCO/NRES	&	2459304.43	&	2021.2435	&	14.20138	&	3.75	&	2.24	&	1.77	&	2.64	\\
460	&	LCO/NRES	&	2459305.30	&	2021.2459	&	14.20181	&	3.56	&	2.13	&	1.65	&	2.64	\\
461	&	LCO/NRES	&	2459322.34	&	2021.2925	&	14.21023	&	3.38	&	2.05	&	1.54	&	2.64	\\
462	&	LCO/NRES	&	2459337.43	&	2021.3338	&	14.21769	&	2.70	&	1.78	&	1.11	&	2.64	\\
463	&	LCO/NRES	&	2459359.38	&	2021.3939	&	14.22854	&	2.73	&	2.01	&	1.17	&	2.64	\\
464	&	LCO/NRES	&	2459379.27	&	2021.4484	&	14.23838	&	2.10	&	1.51	&	1.06	&	2.64	\\
465	&	LCO/NRES	&	2459406.20	&	2021.5221	&	14.25169	&	2.24	&	1.73	&	1.10	&	2.79	\\
466	&	LCO/NRES	&	2459518.85	&	2021.8305	&	14.30738	&	2.13	&	1.54	&	1.20	&	2.79	\\
467	&	LCO/NRES	&	2459543.78	&	2021.8988	&	14.31971	&	2.20	&	1.70	&	1.20	&	2.79	\\
468	&	LCO/NRES	&	2459573.75	&	2021.9808	&	14.33453	&	2.12	&	1.45	&	1.12	&	2.84	\\
469	&	LCO/NRES	&	2459626.65	&	2022.1257	&	14.36068	&	2.30	&	1.48	&	2.34	&	3.03	\\
470	&	LCO/NRES	&	2459706.47	&	2022.3442	&	14.40014	&	2.33	&	1.53	&	2.42	&	3.03	\\
471	&	LCO/NRES	&	2459719.60	&	2022.3801	&	14.40663	&	2.22	&	1.38	&	2.20	&	2.84	\\
472	&	LCO/NRES	&	2459743.50	&	2022.4456	&	14.41845	&	2.31	&	1.50	&	2.42	&	2.84	\\
473	&	LCO/NRES	&	2459750.51	&	2022.4648	&	14.42192	&	2.20	&	1.58	&	2.40	&	2.84	\\
474	&	LCO/NRES	&	2459766.52	&	2022.5086	&	14.42983	&	2.20	&	1.59	&	2.42	&	2.84	\\
475	&	LCO/NRES	&	2459929.83	&	2022.9557	&	14.51057	&	2.30	&	1.50	&	2.40	&	2.82	\\
476	&	LCO/NRES	&	2459932.79	&	2022.9638	&	14.51203	&	2.29	&	1.49	&	2.35	&	2.82	\\
477	&	LCO/NRES	&	2460004.60	&	2023.1604	&	14.54754	&	2.36	&	1.49	&	2.42	&	2.82	\\
478	&	LCO/NRES	&	2460016.81	&	2023.1939	&	14.55357	&	2.50	&	1.47	&	2.50	&	2.82	\\
479	&	LCO/NRES	&	2460030.74	&	2023.2320	&	14.56046	&	2.46	&	1.60	&	2.52	&	2.82	\\
480	&	B. Heatcothe	&	2460026.04	&	2023.2217	&	14.5458	&	2.50	&	1.48	&	2.51	&	2.82	\\
481	&	LCO/NRES	&	2460051.69	&	2023.2894	&	14.57082	&	2.52	&	1.56	&	2.52	&	2.82	\\
482	&	LCO/NRES	&	2460080.63	&	2023.3686	&	14.58512	&	2.45	&	1.58	&	2.58	&	2.82	\\
483	&	LCO/NRES	&	2460094.54	&	2023.4067	&	14.59200	&	2.81	&	1.82	&	3.00	&	2.82	\\
484	&	LCO/NRES	&	2460095.54	&	2023.4094	&	14.59250	&	2.71	&	1.80	&	2.95	&	2.82	\\
485	&	LCO/NRES	&	2460105.57	&	2023.4369	&	14.59745	&	2.30	&	1.51	&	2.39	&	2.82	\\
486	&	LCO/NRES	&	2460113.46	&	2023.4585	&	14.60136	&	2.39	&	1.51	&	2.48	&	2.82	\\
487	&	LCO/NRES	&	2460124.51	&	2023.4887	&	14.60682	&	2.45	&	1.55	&	2.36	&	2.82	\\
488	&	LCO/NRES	&	2460130.25	&	2023.5045	&	14.60966	&	2.69	&	1.53	&	2.40	&	2.82	\\
489	&	LCO/NRES	&	2460154.46	&	2023.5707	&	14.62163	&	2.27	&	1.61	&	2.42	&	2.82	\\
\enddata															
\end{deluxetable}																	

\label{lastpage}
\end{document}